\documentclass[manuscript,nonacm]{acmart}

\AtBeginDocument{%
  }

\setcopyright{none}
\renewcommand\footnotetextcopyrightpermission[1]{}  
\acmConference[]{}{}{} 

\usepackage{array}
\usepackage{longtable}
\usepackage{booktabs}
\usepackage{caption}
\usepackage{tabularx}
\usepackage{tikz}
\usetikzlibrary{shapes.geometric, arrows.meta, positioning, backgrounds, fit}

\begin{document}

\title{Somatic Alignment: Scaffolding Non-Visible Inner Body Sensations in and beyond Vocal Pedagogy}

\author{Ziyue Piao}
\email{ziyue.piao@mail.mcgill.ca}
\orcid{0009-0007-1281-9469}
\affiliation{%
  \institution{IDMIL, MPBL, CIRMMT, McGill University}
  \city{Montreal}
  \state{Quebec}
  \country{Canada}
}

\author{Theodora Nestorova}
\email{tinestorova@viterbo.edu}
\orcid{0000-0003-4224-9368}
\affiliation{%
  \institution{Viterbo University}
  \city{La Crosse}
  \state{Wisconsin}
  \country{USA}
}

\author{Leon Lu}
\email{lv.lu@utoronto.ca}
\orcid{0000-0003-4232-4671}
\affiliation{%
  \institution{University of Toronto Mississauga}
  \city{Mississauga}
  \state{Ontario}
  \country{Canada}
}

\author{Isabelle Cossette}
\email{isabelle.cossette1@mcgill.ca}
\orcid{0009-0005-7899-0192}
\affiliation{%
 \institution{MPBL, CIRMMT, McGill University}
  \city{Montreal}
  \state{Quebec}
  \country{Canada}}

\author{Marcelo M. Wanderley}
\email{marcelo.wanderley@mcgill.ca}
\orcid{0000-0002-9169-4313}
\affiliation{%
  \institution{IDMIL, CIRMMT, McGill University}
  \city{Montreal}
  \state{Quebec}
  \country{Canada}}

\renewcommand{\shortauthors}{Piao et al.}

\begin{abstract}
Inner bodily sensations inform expressive musical performance but can be difficult to interpret and communicate. Through two separate workshops with five singing learners and four senior voice teachers, we investigate breath control in vocal pedagogy. Participants reported difficulty interpreting bodily sensations (\textit{Opacity of Inner Sensations}) and translating awareness into intended adjustments (\textit{Disconnected Regulation of Inner Sensations}), alongside strategies involving touch, demonstration, and imagery. Building on these accounts, we propose the \textit{Somatic Alignment Model (SAM)}, connecting an \textit{Intrapersonal Self-Sensing Loop} with an \textit{Intercorporeal Co-Sensing Loop}. SAM examines how external guidance relates to learners' bodily calibration and how interpretive and regulatory authority is negotiated. We articulate three generative design dimensions—Modality, Agency, and Temporality—and illustrate their possible applications in interactive learning systems and first-person soma design inquiries. SAM offers a conceptual lens for supporting bodily awareness and regulation through interaction design.
 
\end{abstract}




\maketitle

\section{Introduction}

Humans are inherently driven to move, and our ability to perceive, control, and refine complex physical action relies on the seamless integration of sensory-motor feedback loops~\cite{adolph2020importance}. In musical performance, expressive control demands intricate coordination across auditory, tactile, and proprioceptive modalities~\cite{palmer2000conceptual, zimmerman2012multisensory}. However, unlike traditional instrumentalists who interact with visible interfaces offering direct physical resistance, vocalists embody their instrument within the hidden interior of their bodies, making its underlying mechanics non-visible and inherently difficult to control~\cite{leborgne2024vocal}. This difficulty in perceiving internal bodily activity positions singing as a compelling case study for Human-Computer Interaction (HCI): because its mechanisms are deeply interoceptive, performers face persistent challenges in mapping inner somatic sensations to external expressive execution~\cite{leborgne2024vocal, reed2023understanding}.

Central to vocal technique is breath support, a dynamic coordination of multiple muscle groups across the torso that provides the foundational energy for phonation~\cite{leborgne2024vocal, hixon2006respiratory}. Because respiratory activity is not directly visible during singing, vocal instruction often draws on felt bodily cues and spatial metaphors~\cite{hixon2006respiratory, miller1996art}. Learners may therefore find it difficult to relate intended adjustments to what they feel and hear, including when attempts to improve breath support are accompanied by perceived tension or effort. Traditionally, the singing instructor's role is to bridge this divide, guiding students to accurately interpret their own somatic feedback and consciously regulate bodily mechanics, rather than relying solely on imitating the teacher's subjective demonstrations. Bridging this pedagogical divide requires moving beyond one-way demonstration toward real-time co-sensing and mutual bodily feedback~\cite{reed2023negotiating}.

Centuries of vocal pedagogy have cultivated somatic strategies, including hands-on tactile guidance, embodied exercises, and metaphorical verbal description~\cite{leborgne2024vocal}. This work investigates the nuanced practices of vocal pedagogy to explore how non-visible bodily sensations can be shared and calibrated. We derive design considerations grounded in vocal breathing and explore their potential relevance to somatic interaction design more broadly. Our focus is on how external guidance enters a learner's ongoing interpretation and regulation of bodily sensations, and how designers can configure that relationship.

We focus on two research questions:

\begin{itemize}
    \item \textbf{RQ 1:} What difficulties in interpreting inner bodily sensations and regulating breath do singing learners and voice teachers report, and what pedagogical practices do they describe as addressing these difficulties?
    \item \textbf{RQ 2:} How can these accounts inform a conceptual framework for connecting external guidance with a learner's own interpretation and regulation of bodily activity?
\end{itemize}

Synthesizing domain-specific vocal insights with frameworks of soma design~\cite{shusterman2008body, hook2018designing}, metaphor-mediated communication~\cite{reed2023negotiating}, and sensorimotor integration~\cite{wolpert2011principles}, this work offers two core contributions:
\begin{enumerate}
    \item \textbf{Empirical Characterization of Somatic Mismatches and Pedagogical Dynamics:} Qualitative insights from workshops with voice teachers and learners ($N=9$), characterizing two core \textit{Intrapersonal Perception Mismatches} (\textit{Opacity} and \textit{Disconnected Regulation of Inner Sensations}) and identifying the \textit{Intercorporeal Somatic Dynamics} reflected in the multimodal pedagogical practices participants described as addressing these mismatches.
    \item \textbf{The Somatic Alignment Model and Generative Design Space:} A conceptual framework formalizing the dynamics of externalizing and sharing inner bodily sensations by coupling an \textit{Intrapersonal Self-Sensing Loop} with an \textit{Intercorporeal Co-Sensing Loop}. We articulate a three-dimensional generative design space (\textit{Modality}, \textit{Agency}, and \textit{Temporality}) and illustrate how it can inform the reconfiguration of interactive somatic systems and the structuring of first-person soma design inquiries.
\end{enumerate}

\section{Related Work}

To establish a shared vocabulary across music technology, vocal pedagogy, and HCI, key domain-specific terms used throughout this paper are defined in Table~\ref{tab:definitions}.

\begin{table}[htbp]
\caption{Musical and Singing Terminology and Descriptions~\cite{leborgne2024vocal}.}
\label{tab:definitions}
\Description{A two-column glossary table with headers `Terminology' and `Description', organized into three functional categories. Under 'Sound Production', terms include Dynamics, Phonation, and Resonance. Under 'Breath and Support', terms include Appoggio, Breath Support, and Diaphragm. Under 'Vocal Technique', terms include Coloratura, Laryngeal Height, Register, Soft Palate, Straw Phonation, Vocal Fry, and Vocal Tract. Each term is accompanied by a concise definition in the right column.}
\small
\centering
\begin{tabular}{@{} p{0.32\linewidth} p{0.63\linewidth} @{}}
\toprule
\textbf{Terminology} & \textbf{Description} \\
\midrule

\multicolumn{2}{@{}l}{\textit{\textbf{Sound Production}}} \\
\midrule
Dynamics & The variation of loudness in singing (e.g., soft vs. loud), typically controlled through breath pressure and muscular effort. \\
Phonation & The basic sound-making process where vocal folds vibrate to generate voice, initiated and shaped by airflow from the lungs. \\
Resonance & The amplification and coloration of vocal sound through vibration in body cavities (e.g., chest, throat, mouth), similar to how acoustic instruments project sound. \\

\midrule
\multicolumn{2}{@{}l}{\textit{\textbf{Breath and Support}}} \\
\midrule
Appoggio & A classical technique for managing breath support by balancing abdominal and chest muscle engagement to stabilize airflow. \\
Breath Support & The coordinated use of breathing muscles (e.g., diaphragm, ribs, abs) to control airflow and maintain vocal consistency. \\
Diaphragm & A large, dome-shaped muscle under the lungs that helps regulate inhalation and airflow during singing, though it is hard to feel directly. \\

\midrule
\multicolumn{2}{@{}l}{\textit{\textbf{Vocal Technique}}} \\
\midrule
Coloratura & A singing style featuring rapid, ornamented passages (e.g., vocal runs, trills), requiring high breath control and vocal agility. \\
Laryngeal Height & The vertical position of the voice box (larynx), which affects tone and resonance and is adjusted by different singing techniques. \\
Register & A range of vocal pitches (e.g., chest, head voice) produced by different vocal fold vibration patterns; transitions require breath coordination. \\
Soft Palate & The movable part at the back of the mouth's roof; raising it helps create more space for vocal resonance, especially for high notes. \\
Straw Phonation & A vocal exercise using a straw to manage airflow and reduce vocal fold tension, often used in voice therapy. \\
Vocal Fry & A low, creaky vocal sound caused by very relaxed vocal folds and low airflow, sometimes used stylistically but often unintentional. \\
Vocal Tract & The passage from the vocal folds to the lips that shapes the sound of the voice through articulation and resonance adjustments. \\
\bottomrule
\end{tabular}
\end{table}

\subsection{Sensorimotor Integration in Vocal Control}

Music performance represents an intricate sensorimotor task requiring the continuous integration of sensory feedback to coordinate conceptual intentions with fine motor actions~\cite{palmer2000conceptual, crocco2021motor}. Traditional instrumentalists rely extensively on exteroceptive cues (e.g., interface resistance, key strike feedback) paired with proprioception (e.g., limb position, finger trajectory) to calibrate their motor execution in real time~\cite{goebl2008tactile, jensenius2010musical, proske2012proprioceptive}. In contrast, vocal execution functions as an internalized instrument whose mechanics retreat almost entirely into interoception, the sensing of deep internal physiological states such as diaphragmatic descent, subglottic pressure, and visceral core engagement~\cite{hixon2006respiratory, leborgne2024vocal, hook2018designing, bauer2013singing}. 


In vocal pedagogy, instructors rely on a multi-millennial tradition of somatic experiences to communicate these inner body states~\cite{hoch2020historical}, such as physical touch, visual demonstrations, and metaphors (e.g., `breath support', `appoggio') to bridge abstract concepts with tangible bodily sensations~\cite{leborgne2024vocal, reed2023understanding}. These pedagogical practices provide a context for examining how an instructor's additional perspective relates to a learner's interpretation of bodily sensations~\cite{nair1999voice, leborgne2024vocal}.

\subsection{Somatic Mismatches and Alignment}

\subsubsection{Perception Mismatches in Motor Learning}
Motor skill acquisition relies intrinsically on gathering and integrating sensory feedback to guide movement execution~\cite{abbruzzese2003sensorimotor, wolpert2011principles}. In motor control theory, this feedback is mediated by kinesthesia, the internal perception of movement derived from muscular, fascial, and articular receptors~\cite{proske2018kinesthetic}, which the brain integrates into internal predictive models to anticipate sensory outcomes~\cite{kawato2007internal}. When actual kinesthetic feedback diverges from the predicted state, the brain generates a \textit{sensory prediction error}, driving motor adaptation~\cite{shadmehr2010error, krakauer2019motor}.
These accounts explain motor adaptation through differences between predicted and received sensory feedback. Our inquiry addresses a related experiential question: how learners interpret bodily sensations and use them to guide intended adjustments. 

\subsubsection{Intercorporeal Somatic Dynamics}

Pedagogical interactions in somatic disciplines extend beyond isolated, individual sensorimotor loops to encompass a shared, relational space between bodies. Grounded in Merleau-Ponty's phenomenology of intercorporeality~\cite{merleau1964primacy} and Shusterman's somaesthetics~\cite{shusterman2008body}, intercorporeal perspectives foreground how bodily understanding develops through interaction between learners and instructors~\cite{turmo2023intercorporeal}. Within this relational space, we conceptualize somatic dynamics not as rigid postural compliance, but as a reciprocal, co-adaptive feedback loop through which instructor and learner continuously attune to each other's internal bodily states~\cite{kleinman2013alexander}. Cultivating this attunement is vital: because prescriptive physical corrections often trigger muscular tension, true sensorimotor learning depends on guided self-regulation and shared bodily awareness. Remarkably, this pursuit of somatic alignment transcends temporal and cultural boundaries, resonating across ancient wisdom and modern embodied practices alike, from Daoist bodily cultivation and Yogic traditions to Western somatic disciplines such as the Alexander Technique and Feldenkrais Method~\cite{kleinman2013alexander}. Across these diverse domains, cultivating heightened bodily awareness and intersubjective co-sensing emerges as a foundational prerequisite for mastering tacit motor skills.

In practice, instructors utilize subtle, non-coercive touch to help learners release tension and co-experience proper alignment~\cite{gelb2013body}. In singing instruction, Alexander Technique practitioners pair this tactile guidance with continuous verbal co-inquiry (e.g., prompting: \textit{``Is your spine flexible and moving as you breathe?''}) to calibrate mutual understanding in real time~\cite{kleinman2013alexander, sentianin2018natural}. 
However, even when instructors accurately perceive these mechanics, they must translate such hidden bodily processes into localized gestures, tactile prompts, and evocative metaphors. Vocal training has continually refined these embodied methods to negotiate the difficulty of perceiving internal bodily activity, yielding a remarkably rich repertoire of multimodal interactions that offers invaluable, transferable design wisdom for Soma Design.

\subsection{Embodied Interaction and Externalizing Inner Sensations}

While classical physiology strictly separates internal sensing into \textit{interoception}~\cite{craig2002you} and \textit{proprioception}~\cite{proske2012proprioceptive}, such rigid binaries fail to capture the lived reality of vocal performance. Singers experience these pathways as an integrated somatic continuum bridging physical mechanics, cognitive focus, and emotional expression. Drawing on the first-person ethos of Soma Design~\cite{hook2018designing}, we use \textit{inner sensations} as a holistic construct for the first-person experience of internal mechanics, a felt feedback loop through which performers monitor, regulate, and externalize bodily states during motor skill acquisition.

Research into intra- and intercorporeal dynamics~\cite{patibanda2024exploring} has drawn growing interest from HCI researchers, particularly in augmenting human motor capabilities and facilitating shared bodily control. Studies demonstrate that negotiating somatic states is essential for establishing intercorporeal understanding and fostering effective collaboration~\cite{eriksson2019dancing, kilic2023corsetto, Stepanova2024intercorporeal, turmo2023intercorporeal}. Specifically, Reed et al.~\cite{reed2023negotiating} proposed a \textit{Model of Metaphor-Based Communication}, which formalizes how subjective bodily knowledge is shared between a communicating agent (e.g., a teacher) and a receiving agent (e.g., a student). In this framework, lived experience is encoded into multimodal metaphors and subsequently internalized through an iterative loop that leverages ambiguity to establish mutual understanding. Building on accounts of metaphor-mediated communication and intercorporeal biofeedback, SAM focuses on how an additional perspective enters a learner's ongoing regulation: how a cue is interpreted, how an adjustment is attempted, and how its felt consequences inform subsequent calibration. Its intended contribution is a vocabulary for relating these processes and examining the distribution of interpretive and regulatory authority.

\section{Methodology}

\begin{table*}[htbp] 
\centering
\renewcommand{\arraystretch}{1.3} 
\caption{Participant Demographics and Profile Information. All profiles are based on participants' self-reports.}
\label{participants}
\Description{A demographic table structured into two main sections across five columns: ID, gender, experience, profile, and format. The first section lists five student participants (S1 through S5), consisting of three women and two men who attended onsite. Their singing experience spans from self-taught to over 20 years across pop, choir, jazz, musical theatre, and classical traditions. The second section lists four vocal teachers (T1 through T4), all women who participated online. Their teaching experience ranges from over 10 to over 29 years, covering cross-genre vocal instruction, clinical vocal rehabilitation, and diverse age groups from youth to older adults.}
\begin{tabular}{>{\raggedright\arraybackslash}p{0.8cm} >{\raggedright\arraybackslash}p{1.2cm} >{\raggedright\arraybackslash}p{2.3cm} >{\raggedright\arraybackslash}p{8.5cm} >{\raggedright\arraybackslash}p{1.5cm}}  
\toprule

\multicolumn{5}{c}{\textbf{Student Workshop ($N=5$)}} \\ \midrule
\textbf{ID} & \textbf{Gender} & \textbf{Learning Exp.} & \textbf{Student Profile} & \textbf{Format} \\ \midrule
S1 & Woman & Over 14 years & Primarily choir experience with two years of musical theatre. & Onsite \\ \hline
S2 & Woman & Over 3 years  & Trained in Western Classical music for Art Songs and Operas. & Onsite \\ \hline
S3 & Woman & Over 20 years & Background in Indian Classical music, Pop, and Jazz. & Onsite \\ \hline
S4 & Man   & Self-taught only   & Focused on Pop music. & Onsite \\ \hline
S5 & Man   & Over 1 year   & Experience in Pop and Western Classical music. & Onsite \\ 

\midrule 
\multicolumn{5}{c}{\textbf{Teacher Workshop ($N=4$)}} \\ \midrule
\textbf{ID} & \textbf{Gender} & \textbf{Teaching Exp.} & \textbf{Teacher Profile} & \textbf{Format} \\ \midrule
T1 & Woman & Over 29 years & Teaches vocal techniques across genres. Works with trans women, cisgender professionals, amateurs (primarily adults), and teenagers in preschool to middle school choir. & Online \\ \hline
T2 & Woman & Over 20 years & Teaches Classical international styles (Bollywood, South African, Sudanese), Musical Theatre, Rock and Roll, and vocal rehabilitation. Works mainly with university students and vocal clinical cases. & Online \\ \hline
T3 & Woman & Over 10 years & Teaches Western Classical, Musical Theatre, Pop, and Folk. Works with university students, children, and older adults. & Online \\ \hline
T4 & Woman & Over 15 years & Teaches Western Classical, Musical Theatre, and Pop. Works with young professionals, postgraduate students, pre-university students, amateur choral singers, and retired singers focusing on vocal health. & Online \\ 
\bottomrule
\end{tabular}
\end{table*}

To investigate the somatic mismatches and pedagogical practices in vocal instruction and derive design implications for closed-loop somatic systems, we conducted two workshops involving focus group discussions and body mapping sessions with singing learners and voice professors. To preserve role-specific perspectives, participants were grouped by role (students or teachers).
We conducted the first workshop with students from diverse genres and levels and another workshop with expert vocal instructors with multi-genre teaching experiences. Participant demographics are summarized in Table~\ref{participants}, with students designated as ``S[number]'' and teachers as ``T[number]''.


\subsection{Participant Recruitment}

This study was approved by the Institutional Review Board at McGill University. To engage experienced voice educators, recruitment was conducted globally through international professional networks (e.g., The Voice Foundation and Pan American Vocology Association), while student singers were recruited via academic mailing lists and postings at a university music department and an affiliated music research center. A screening survey assessed prospective participants' training background, performance genres, and teaching tenures to ensure rich domain expertise and pedagogical diversity. 

In total, nine participants were recruited, comprising four senior voice pedagogues and five singing students who engaged in in-depth somatic workshops. The instructor cohort comprised four women who reported teaching experience ranging from over 10 to over 29 years across genres, including classical, contemporary commercial, and musical theatre. The student cohort also spanned varied gender identities and musical traditions. Our sampling rationale drew on information power, emphasizing the focused study aim and participants' experience relevant to vocal breath pedagogy~\cite{malterud2016sample}. All participants provided written informed consent at the beginning of each workshop. 

Critically, instructors and students were separated into distinct workshop cohorts to mitigate master-apprentice power dynamics common in vocal training. Separating the groups was intended to reduce evaluative pressure and encourage candid discussion.
Sessions were conducted in either online or in-person formats using identical protocols. In both formats, we utilized Mural~\footnote{https://www.mural.co/, a collaborative digital whiteboard} for brainstorming and note-taking. For the body mapping sessions, onsite participants used physical paper templates, while online participants used digital canvases in Mural. Participants were compensated for their time according to local institutional standards.

\subsection{Workshop Process}

\begin{table*}[t]
\caption{Chronological Focus Group Protocol and Alignment with Research Questions.}
\label{tab:procedure}
\Description{A four-column table outlining the seven chronological phases of the workshop protocol across phase number, activity format, activity focus, and prompt directives. Phases 1 through 3 consist of focus group discussions covering participant introductions, brainstorming breath challenges, and identifying core perceptual mismatches. Phase 4 transitions to an individual body mapping exercise marking torso areas of disconnect on student templates. Phase 5 uses a focus group format to examine pedagogical interventions, while Phases 6 and 7 conclude with body mapping activities exploring intercorporeal feedback across teacher-student outlines and co-envisioning future wearable interfaces.}
\small
\centering

\renewcommand{\arraystretch}{1.5} 
\begin{tabularx}{\textwidth}{@{} c >{\raggedright\arraybackslash}l >{\raggedright\arraybackslash}p{0.34\textwidth} >{\raggedright\arraybackslash}X @{}}
\toprule
\textbf{Phase} & \textbf{Activity Format} & \textbf{Activity Focus} & \textbf{Prompt Directive} \\
\midrule
1 & Focus Group & Demographic Introduction & ``Can you briefly introduce your singing learning or teaching background, primary genres, and years of practice?'' \\
2 & Focus Group & Brainstorming Respiratory Challenges & ``What physical challenges do you (or your students) encounter with breath control during singing? Describe the internal bodily sensations experienced.'' \\
3 & Focus Group & Identifying Core Perceptual Mismatches & ``Which challenge represents the most significant disconnect between inner sensation and actual physical execution in your singing (or your students' practice)?'' \\
4 & Body Mapping & Intrapersonal Respiratory Disconnects & ``Which body areas feel (or are observed to be) most disconnected, tense, or confusing when managing breath? Mark them on the \textit{Student} template.'' \\
5 & Focus Group & Exploring Pedagogical Interventions & ``What pedagogical interactions, cues, or exercises have helped align inner breath sensations with instructional intent?'' \\
6 & Body Mapping & Intercorporeal Alignment \& Feedback & ``Using the \textit{Student} and \textit{Teacher} outlines, draw how breath feedback is physically transmitted, felt, and interpreted across both bodies.'' \\
7 & Body Mapping & Envisioning Wearable Systems & ``What next-generation technologies could help resolve these breathing mismatches? Draw your imagination on the body template.'' \\
\bottomrule
\end{tabularx}
\end{table*}

Each 90-minute workshop alternated between four focus group discussion sessions and three body mapping sessions, as detailed in Table~\ref{tab:procedure}. For the discussion sessions, we selected focus groups over individual interviews or classroom observations to foster collaborative reflection, cross-genre dialogue, and the distillation of shared kinesthetic experiences across diverse training levels and experiences within the same group of people. To record inner bodily sensations, we employed \textit{body mapping}, a soma-based design method in which participants visually externalize physical sensations, muscle engagement, and perceptual disconnects on body outlines~\cite{anne2022body}. 
To ensure methodological rigor, the workshop protocol was iteratively refined through a preliminary pilot session with two singing learners with different levels of experience, alongside weekly meetings between the first three authors, drawing upon their combined backgrounds in design, vocal performance, and voice pedagogy. Table~\ref{tab:procedure} lists the seven protocol phases.

During \textit{Focus Group Discussions}, participants posted sticky notes within dedicated Mural zones. During \textit{Body Mapping} sessions, participants individually annotated paper or digital torso outlines to locate non-visible respiratory mechanics and areas of tension. 

\begin{figure*}[t]
  \centering
  \includegraphics[width=\linewidth]{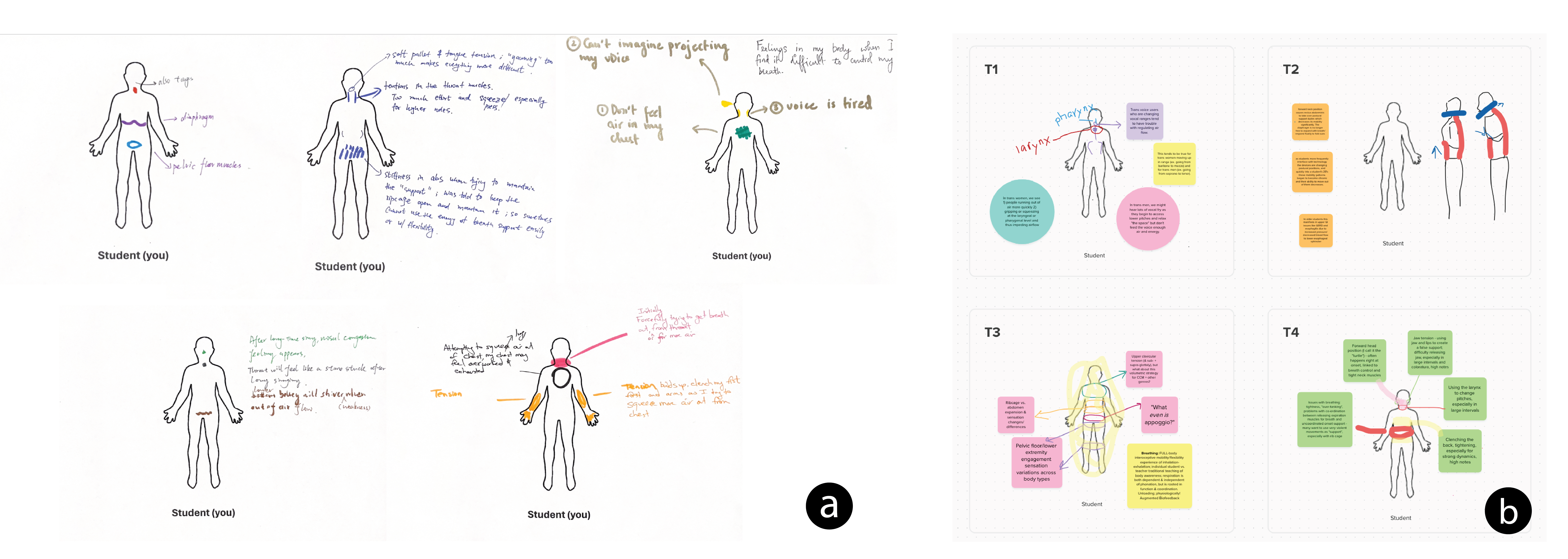}
  \caption{Representative body mapping artifacts highlighting non-visible respiratory sensations and tension patterns across (a) in-person paper templates and (b) online digital canvases. Markings illustrate intrapersonal disconnects (e.g., diaphragmatic opacity, clavicular tension) and intercorporeal feedback loops, which directly inform design requirements for multi-site wearable respiratory interfaces.}
  \Description{A two-part visual comparison of annotated body maps. On the left, panel (a) contains five scanned paper templates showing outlined human bodies titled 'Student (you)'. Hand-drawn markings, shaded color patches, and handwritten notes point to localized sensations across the throat, chest, and lower abdomen. On the right, panel (b) displays four digital whiteboard workspaces labeled T1 through T4. Each workspace features a stylized human silhouette annotated with colored rectangular sticky notes, posture diagrams, and callout arrows pointing to the diaphragm, spine alignment, and respiratory core.}
  \label{fig:bodymapping}
\end{figure*}

\subsection{Data Analysis}

We compiled a multimodal dataset comprising transcribed audio recordings, collaborative whiteboard notes, and annotated body maps. Reflexive Thematic Analysis (RTA)~\cite{braun2021one} was chosen for its capacity to integrate diverse qualitative media beyond textual transcripts and for valuing researcher reflexivity in interpreting nuanced, embodied experiences. The collaborative research team consisted of the first author (a semi-professional classical singer and design researcher), the second author (a senior voice pedagogy teacher), and the third author (an experienced design researcher).
Guided by the research questions, our analytical pipeline proceeded as follows:
\begin{enumerate}
    \item \textbf{Data Familiarization \& Transcription:} Audio recordings were transcribed using NVivo~\cite{NVivo} and manually verified by the first author. Transcripts were read iteratively alongside body map annotations to construct a holistic, contextualized understanding of each session.

    \item \textbf{Multimodal Open Coding:} The first author conducted initial open coding on transcripts while concurrently mapping visual annotations by anatomical region (e.g., lower abdomen, ribcage, clavicle, throat) and felt sensation types (e.g., tension, vacancy, support, directional airflow) as articulated by participants.

    \item \textbf{Thematic Generation \& Collaborative Reflexivity:} Through an iterative process, initial codes were clustered into candidate themes corresponding to our research questions. Weekly collaborative meetings were held with the research team to reflexively interrogate assumptions, explore interpretive tensions, and richly describe contrasting student and instructor perspectives.

    \item \textbf{Domain-Grounded Reflexive Tuning:} The team engaged with the synthesized themes to ensure that the generated themes meaningfully bridged pedagogical realities with actionable design requirements for wearable interfaces.
\end{enumerate}

\section{Findings}
\label{sec:findings}

Our analysis reveals a contrast in communication styles: while students relied heavily on body maps paired with verbal descriptions to articulate complex internal sensations, voice teachers communicated primarily through established somatic metaphors and physiological models, using the visual maps more sparingly. Consequently, the body maps presented as figures in this section predominantly illustrate student perspectives.
Through RTA, we identified three primary thematic domains: (1) \textit{Intrapersonal Perception Mismatches}, (2) \textit{Intercorporeal Somatic Dynamics}, and (3) \textit{Co-Envisioning Future Interfaces}. 
\subsection{Intrapersonal Perception Mismatches}
\label{sec:findings_intrapersonal}

This theme examines \textit{Intrapersonal Perception Mismatches}: reported difficulties in interpreting bodily sensations and relating intended adjustments to perceived bodily and vocal outcomes. Specifically, we identify two primary dimensions of such mismatch: \textit{Opacity of Inner Sensations} and \textit{Disconnected Regulation of Inner Sensations}.

\subsubsection{Opacity of Inner Sensations}
\label{sec:OIS}

\begin{figure}[h!]
  \centering
  \includegraphics[width=0.5\linewidth]{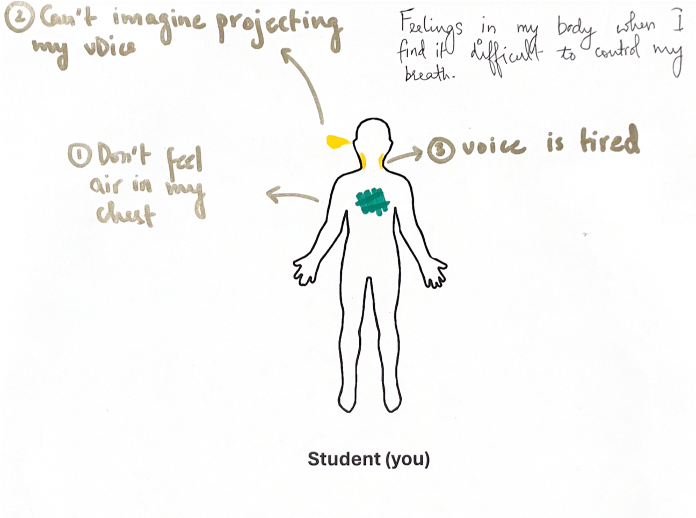}
  \caption{Illustration of a student's self-reported challenges with vocal projection and breath control, depicting physical sensations such as a tired voice, difficulty breathing, and a lack of air in the chest during performance.}
  \Description{A hand-drawn body map on a template labeled 'Student (you)', featuring callout arrows and colored highlights on a minimalist human outline. A dark green scribble covers the center of the chest, accompanied by an arrow pointing to note 1: 'Don't feel air in my chest'. Yellow shading marks the throat, labeled note 3: 'voice is tired'. Another yellow mark sits beside the ear, labeled note 2: 'Can't imagine projecting my voice'. In the top right corner, handwritten text notes: 'Feelings in my body when I find it difficult to control my breath'.}
  \label{fig:voicetired}
\end{figure}

A fundamental barrier to embodied learning is the \textit{Opacity of Inner Sensations}. Participants described difficulty locating and interpreting sensations associated with internal bodily activity, sometimes relying on vague or poorly localized cues (e.g., S1, gesturing vaguely toward their abdomen: \textit{``I feel a little bit sour here''}). Instructors emphasized that making these invisible internal mechanics visible is critical. T1 described perceived improvements in breathing efficiency and sound quality following clarification of anatomical misconceptions, noting: \textit{``after watching videos of someone misunderstanding where the diaphragm is or how breath moves during inhalation, I was amazed at how much of a difference it made to the efficiency of their breathing and the quality of their sound.''}

To compensate for this opacity, pedagogy often relies on abstract conceptual labels (e.g., \textit{``support''} or \textit{appoggio}) to describe holistic motor coordination. However, because internal somatic awareness is inherently subjective, learners often struggle to translate static verbal instructions into unified physical actions, experiencing a sense of somatic disconnection between intent and execution. S2 articulated this spatial disconnect between abstract instruction and lived bodily feeling: \textit{``I find it difficult to harness breath support effectively... I don't really feel the energy here [pointing to abdomen]. There's a disconnect between the upper and lower parts of my body.''} Instructors acknowledged that static verbal cues struggle to convey tacit bodily knowledge because, as T3 observed: \textit{``Breathing involves a full-body interoceptive experience... but it varies between individuals.''}

Participants also described familiar breathing sensations and strategies becoming harder to access during changes in bodily condition. For example, S1 described difficulty controlling breath and engaging the abdominal and pelvic-floor regions during menstruation: \textit{``I find it very difficult to control my breath during my period... it's definitely hard to engage the muscles here [pointing to the abdomen and pelvic floor area].''} T1 similarly recalled difficulty breathing and producing a focused sound after pregnancy and a Cesarean section: \textit{``After pregnancy, breathing felt nearly impossible, and after my first C-section, I couldn't produce a focused sound.''} S1 also reported changes in breath preparation and vocal stability following COVID: \textit{``After COVID, if I'm not well-prepared with breath support, my sound breaks into pieces.''}

Together, these accounts suggest that difficulties interpreting inner sensations involve unclear bodily reference points, ambiguous instructional language, and changes in familiar bodily experience. They motivate exploring whether external representations could help learners interpret these experiences.

\subsubsection{Disconnected Regulation of Inner Sensations}
\label{sec:DRIS}

\begin{figure*}[h!]
  \centering
  \includegraphics[width=\linewidth]{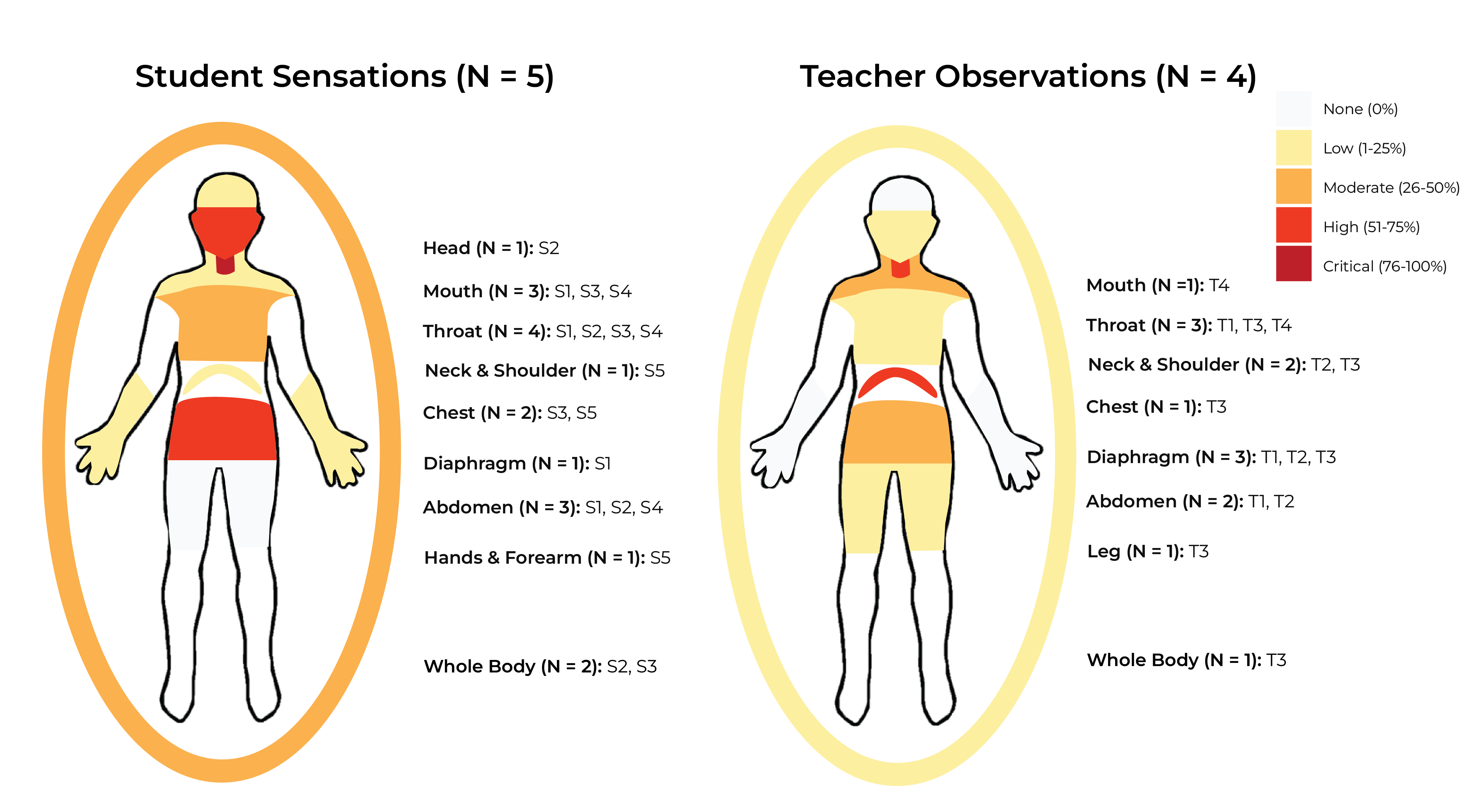}
  \caption{Comparative somatic heat maps depicting the anatomical distribution of respiratory mismatches. Color intensity represents the normalized frequency of student self-reported sensations ($N=5$, left) versus teacher-observed pedagogical focal points ($N=4$, right).}
  \label{fig:mismatchheatmap}
  \Description{Two side-by-side stylized human body silhouettes displaying heat maps, with a five-tier color scale in the upper right ranging from light yellow (Low, 1–25\%) to dark red (Critical, 76–100\%). On the left silhouette, titled 'Student Sensations (N=5)', high color intensity is concentrated in the upper body and abdomen, featuring solid dark red across the throat (N=4) and lower abdomen (N=3), alongside orange and yellow bands across the chest (N=2), mouth (N=3), neck and shoulders (N=1), and head (N=1). On the right silhouette, titled 'Teacher Observations (N=4)', the primary focal point shifts to a sharp, dark red curved band highlighting the diaphragm (N=3), accompanied by moderate orange shading on the throat (N=3) and lighter yellow shading across the abdomen (N=2), neck and shoulders (N=2), chest (N=1), and leg (N=1).}
\end{figure*}

While \textit{Opacity} concerns difficulty interpreting bodily sensations, \textit{Disconnected Regulation} concerns reported difficulty translating awareness of an unwanted sensation or vocal outcome into an adjustment that feels effective. The somatic heat maps (Figure~\ref{fig:mismatchheatmap}) show different anatomical emphases in the two groups' accounts: students emphasized discomfort surfaces like the throat, mouth, and chest, whereas teachers emphasized missing support in deeper postural regions such as the abdomen and lower torso.

Teachers described some learners' efforts to support demanding vocal passages as involving unwanted tension in the jaw, lips, or throat. T4 identified this as an involuntary reliance on false motor support: \textit{``Students often use their jaws and lips to create false support, which makes it difficult to relax, especially during large intervals, coloratura passages, and high notes.''} This misdirected regulation distorts the practitioner's tactile experience. S4 described navigating high registers as \textit{``a state of intense physical conflict, feeling like squeezing and crashing with a throat that turns like a stone.''} Participants associated tension in these regions with difficulties managing breath. As S5 observed, \textit{``constriction emerges from forcefully trying to push air out of the throat,''} while S1 corroborated that \textit{``when I lose control of my breath, tension builds in the vocal tract, and my tongue becomes uncooperative.''}

Participants also described contrasting sensations in the torso, including excessive effort and a sense of insufficient support. S5 reported that \textit{``trying to force air from my chest makes it feel overworked and exhausted,''} while S3 experienced the opposite sensation of a collapsing somatic frame: \textit{``my chest feels empty, and I struggle to project my voice; it feels like it's being pulled inward.''} T4 associated some attempts at muscular engagement with tension in the back, noting: \textit{``improper muscular engagement routinely induces back tension, particularly during high-intensity dynamics.''}

Even when learners attempt to consciously regulate the correct core area, their efforts sometimes result in over-correction and a sense of stiffness during attempted adjustment. Teachers described some learners' attempts to maintain support as overly effortful or rigid. T4 highlighted this regulatory error: \textit{``Some students try to keep their stomach flat, which limits their ability to breathe and release tension. I tell them to engage only about 10\% of the muscles they think are necessary.''} S2 described deliberate attempts to maintain abdominal support as producing stiffness: \textit{``When I try to maintain support in the abdomen, everything feels stiff and wrong.''} T2 interpreted similar difficulties in terms of restricted diaphragmatic movement: \textit{``the diaphragm becomes restricted, unable to expand or respond fluidly to breath cues.''}

Because these disconnected regulatory patterns solidify into deeply ingrained motor reflexes, they form a self-reinforcing cycle of physical fatigue and control breakdown. S3 also described vocal fatigue occurring alongside difficulty controlling breath: \textit{``Sometimes, my voice feels very tired, and I lose control of my breath.''} Together, these accounts characterize difficulty translating bodily awareness into adjustments that feel effective, sometimes alongside unwanted effort or fatigue.


\subsection{Intercorporeal Somatic Dynamics}
\label{sec:intercorporeal}
To mitigate the perceptual opacity and regulatory disconnects inherent in vocal breathing, pedagogical practice relies on dynamic bodily strategies that extend beyond static verbal instruction. This subsection examines \textit{Intercorporeal Somatic Dynamics}, the interactive mechanisms through which teachers and learners co-construct, externalize, and align bodily awareness. We delineate these practices across two complementary layers: first, how instructors scaffold a student's individual somatic awareness through movement practices and tangible props; and second, how they bridge subjective sensory barriers through direct intercorporeal touch and shared metaphorical imagery.

\subsubsection{Supporting Students' Somatic Awareness}
Teachers use physical practices and tangible tools to help learners perceive and control their own internal mechanics. Rather than relying solely on sound, these exercises augment a student's inner sensations, providing physical constraints or dynamic counterweights that allow them to perceive and regulate their own body. 

To decouple breath support from reflexive skeletal bracing, instructors described introducing somatic interventions adapted from yoga, Pilates, and athletic conditioning. T4 illustrated how dynamic balance exercises disrupt static tension, allowing learners to access core support in motion: \textit{``I incorporate exercises like standing on one leg while swinging the other to explore support without engaging tense muscles... I also use yoga-inspired poses to engage and release muscles simultaneously.''} Instructors described pairing these whole-body movement scaffolds with continuous vocalises, such as siren glides and vowel appoggios, to reinforce flexible coordination across registers. Furthermore, T1 incorporated rehabilitative physical therapy to deconstruct habitual tension, noting that while targeted somatic interventions alleviate constriction, internal calibration remains an iterative process: \textit{``Pilates-based rehabilitation... significantly helped release the muscle tension. Techniques like the Mendelsohn maneuver are helpful, but finding the right laryngeal height takes trial and error.''}

Beyond dynamic movement routines, instructors regularly deploy tangible artifacts and physical props to convert elusive interoceptive cues into tangible somatosensory feedback. Tactile and resistive tools, such as scarves and elastic exercise bands, are frequently used to provide physical anchoring against the torso, helping learners perceive core expansion without tensing peripheral muscle groups. As T4 explained: \textit{``I use physical tools like elastic bands and scarves to help students feel free support in the body... [these tools] help students understand support without unnecessary tension, which can particularly free the back and improve ease in the upper register.''} Similarly, semi-occluded exercises such as straw phonation introduce tangible back-pressure at the lips, transforming intangible internal breath flow into direct tactile resistance that guides dynamic airflow regulation. However, instructors acknowledged the fundamental limitations of these passive, low-tech props, which still rely on trial-and-error interpretation and constant pedagogical oversight. As T2 reflected: \textit{``Sometimes, all we can do is to point things out, demonstrate, and hope they internalize it. But it's challenging because I've never seen a student perfectly aligned at the start of a lesson.''}

\subsubsection{Intercorporeal Sensing and Communication with Somatic Prompts}

\begin{figure}[h!]
  \centering
  \includegraphics[width=0.5\linewidth]{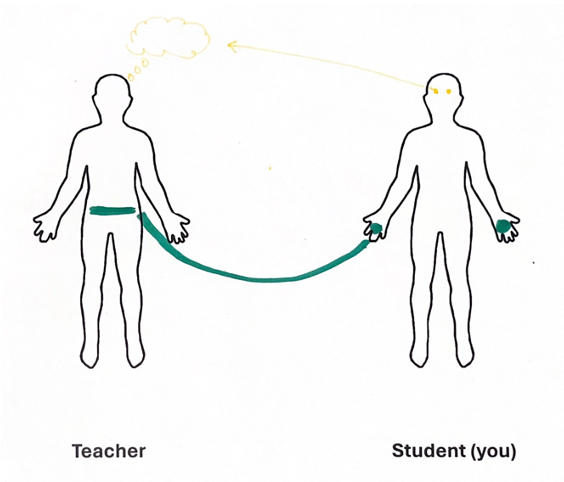}
  \caption{Illustration from S3 of a hands-on teaching method where the student touches the teacher's abdomen to feel muscle movement and understand breath support.}
  \label{fig:touch}
  \Description{A minimalist line drawing showing two human figures standing side by side, labeled 'Teacher' on the left and 'Student (you)' on the right. The teacher figure features a wide green band across the lower abdomen, from which a curved green line loops downward to connect to both hands of the student, which are highlighted in green. Above their heads, a faint yellow dashed line with a thought bubble connects the teacher to the student.}
\end{figure}

When explicit instructions targeting specific movements fail to convey hidden internal mechanics, teachers and students described \textit{intercorporeal communication}, transitioning from concrete physical adjustments to abstract somatic sensations, imagery, and bodily metaphors to externalize and align subjective states.

A primary conduit for this somatic transfer is intercorporeal physical touch (Figure~\ref{fig:touch}), which grounds hidden physiological mechanics through direct haptic transmission. Instructors frequently invite students to palpate the teacher's lower abdomen or dorsal region to perceive dynamic muscular expansion and engagement during phonation. S2 emphasized how this direct tactile contact demystified abstract breathing mechanics: \textit{``For support, my teacher had me touch her lower belly to feel how the muscles work with breath, moving inward or outward. This helped me understand the mechanism in the body.''} For learners, experiencing this physical co-presence often constituted a pivotal breakthrough in their vocal training; as S1 recounted, sensing the instructor's physical coordination was \textit{``the first time I really learned how to breathe for singing... my teacher let me touch her muscles and had me imitate that to feel it myself.''} T4 likewise described direct tactile modeling as more educational than reading about technique, reflecting: \textit{``When students touch me to feel coordination, it's far more educational than reading about technique... and when I did [experience it], it was much more impactful.''}

\begin{figure}[h!]
  \centering
  \includegraphics[width=0.5\linewidth]{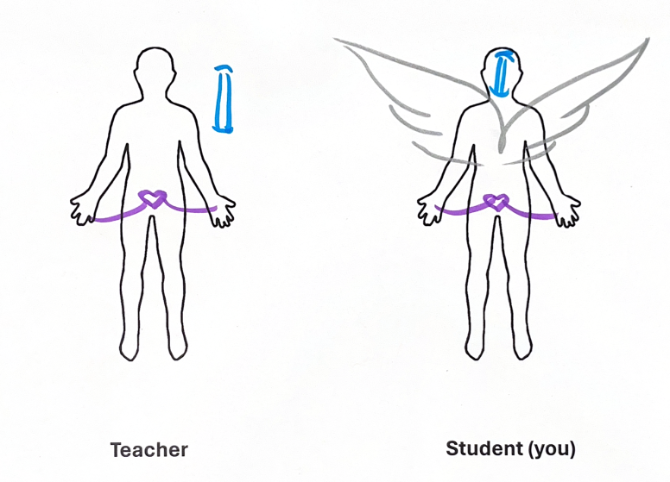}
  \caption{Illustration from S1 of a teacher and student using mental imagery in singing instruction.}
  \label{fig:metaphor}
  \Description{A minimalist line drawing showing two human outlines standing side by side, labeled 'Teacher' on the left and 'Student (you)' on the right. Both figures feature a purple ribbon spanning horizontally across their hands with a heart-shaped knot at the lower abdomen. Floating beside the teacher's head is a blue vertical double-ended arrow. The student figure is depicted with large sketched wings extending from the upper back and shoulders, along with a matching blue vertical double-ended arrow drawn inside the head and neck.}
\end{figure}

Participants described imagery as a way to connect verbal guidance with felt qualities of movement. S5 highlighted this reliance on imaginative translation: \textit{``I receive the teacher's advice and try to imagine it in my mind.''} By anchoring abstract physical instructions to vivid mental models, these metaphors facilitate a cross-modal integration that bridges \textit{``the visual image with thinking and feeling''} (S2), enabling more organic motor coordination.
Participants described spatial and movement-based imagery as helpful for approaching demanding vocal tasks with less perceived tension and a greater sense of ease (Figure~\ref{fig:metaphor}). S1 recalled using expansive imagery to navigate the upper register without constriction: \textit{``My teacher once told me to imagine I have wings lifting me... that imagery helped me feel a sense of flight and energy in my singing.''} Conversely, counter-intuitive physical imagery helped disarm overactive preparation; S2 associated imagining a sensation of ``falling during singing'' with a sense of release and openness. Furthermore, spatial reframing enabled singers to decouple pitch height from muscular effort. As S3 explained: \textit{``My teacher helped me break the mental tension around high and low notes. Instead of thinking of them vertically, she encouraged me to view them horizontally, which shifted my perception and relaxed my approach to higher pitches.''} Ultimately, internalizing these metaphorical mappings into effortless somatic execution remains \textit{``an iterative process of somatic exploration, requiring learners to [overdo] it until the right feeling is found''} (S2).


\subsection{Co-Envisioning Future Interfaces}
\label{subsec:findings-tech}

Looking beyond traditional vocal pedagogy, participants co-envisioned interactive technologies to help bridge these perceptual disconnects. However, their visions exposed a central dilemma: while singers intuitively sought direct external cues like visual displays and haptic feedback, participants cautioned that relying on rigid automated feedback risks eroding a performer's intuitive bodily awareness.

\subsubsection{Externalizing Deep Anatomy: Visual Guidance vs. Internal Awareness}

To mitigate the \textit{Opacity of Inner Sensations}, participants primarily envisioned dynamic anatomical visualizations to render hidden physiological mechanics concrete and actionable. As S3 noted, \textit{``seeing [internal anatomy] makes understanding easier,''} underscoring how visual feedback is instinctively sought to bridge the chasm between subjective perception and physical reality.

Both cohorts prioritized visualizing somatic mismatch zones (Figure~\ref{fig:mismatchheatmap}). T4 emphasized that \textit{``being able to see how the diaphragm works during breathing makes a huge difference,''} while T2 envisioned dynamic spatial models to cultivate spatial mapping: \textit{``I once saw a kinetic ball expand with a breath... If students could interact with detailed, 3D anatomy visuals [of the diaphragm], it would help them form a mental map and guide their body movement.''} In the upper tract, S4 proposed cross-sectional imaging (e.g., an \textit{``X-ray-like image [showing] tongue placement during singing''}) to clarify phoneme execution.

This demand extended toward comparative intercorporeal modeling, where learners desired to compare their internal physiology side-by-side with the instructor's. S4 expressed wanting \textit{``my teacher to help visualize which muscles work and how, so I can mimic accurately,''} and S1 desired to \textit{``see a cross-section of my teacher's body and mine to understand how to position my tongue and mouth.''}

Although participants readily welcomed visual tools, this tendency reflects a potential challenge in embodied learning, namely, approaching internal bodily processes through external imitation alone. As instructors pointed out, an overemphasis on visual feedback risks shifting focus outward, which may inadvertently hinder the learner's ability to cultivate their own interoceptive awareness. As T2 reflected, \textit{``students may become too dependent on visual aids, ignoring the bodily sensations necessary for healthy singing.''} Moreover, instructors noted that because breathing mechanics vary widely across morphologies, visual displays must dynamically accommodate individual bodily differences rather than enforcing a singular, standardized anatomical template. Rather than relying on literal, photographic realism, visual interfaces could instead embody dynamic, evocative metaphors to guide spatial awareness without inducing cognitive fixation.

\subsubsection{Tactile Interventions: Corrective Prompts vs. Somatic Scaffolding}

\begin{figure}[h!]
  \centering
  \includegraphics[width=0.4\linewidth]{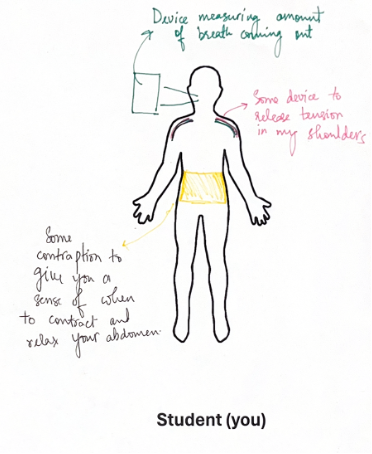}
  \caption{Illustration of S3's vision for tactile feedback devices to support singing.}
  \Description{A hand-drawn sketch of a standing human figure annotated with three envisioned assistive devices. To the upper left, a green rectangular box is drawn beside the mouth with an arrow pointing to the text: 'Device measuring amount of breath coming out'. Arched colored stripes in pink, purple, and blue rest over both shoulders, with a pink arrow pointing to the text: 'Some device to release tension in my shoulders'. Across the waist and lower abdomen is a wide yellow-shaded band, with a yellow arrow pointing to the note: 'Some contraption to give you a sense of when to contract and relax your abdomen'.}
  \label{fig:techtouch}
\end{figure}

To mitigate the \textit{Disconnected Regulation of Inner Sensations}, participants envisioned wearable tactile devices offering closed-loop feedback. Faced with sensory ambiguity, learners initially asked for direct corrective prompts in high-mismatch torso areas (Figures~\ref{fig:mismatchheatmap} and \ref{fig:techtouch}). S1 sought cues that \textit{``indicate exactly when to contract or relax abdomen muscles,''} while S5 suggested distributed vibration where \textit{``a buzzer or similar device placed on different parts of the body could signal exactly where muscle engagement is required.''}

When addressing compensatory strain, participants shifted toward passive tension awareness and release. S3 wanted technology to interrupt chronic holding habits: \textit{``A device measuring shoulder tension and showing how to release it would significantly reduce accumulated tension.''} S4 similarly noted that tactile tools should clarify confusion rather than command muscle action: \textit{``I often wonder which muscles should be tense or relaxed... A device that could visually or tactilely indicate muscle use, particularly around my face, throat, and abdomen, would be incredibly helpful.''}

While learners intuitively ask for haptic devices like buzzers to automate timing, prescriptive haptic cues risk triggering unintended muscular bracing and compensatory tension, as observed in prior evaluations of haptic singing guidance systems~\cite{piao2022sensing}. Taken together, these accounts and prior findings motivate exploring haptic cues as experiential reference points for self-inquiry, alongside their possible use for explicit guidance. To achieve this in authentic studio practice, instructors emphasized that hardware must avoid intrusive form factors and cumbersome setup rituals that disrupt pedagogical flow. As T2 noted, technical friction quickly breaks lesson momentum: \textit{``setup takes so much time. Ideally, it should be as simple as `stick on, machine on, go,' but it often takes 20 minutes just to get started.''} Consequently, T3 advocated transitioning away from rigid laboratory equipment toward studio-ready form factors, such as \textit{``respiratory belts, but more user-friendly and designed for in-studio use... allow[ing] for comparisons across body types and perception differences.''} Wearable somatic interfaces must therefore offer lightweight, seamless integration that respects bodily freedom and accommodates individual morphological diversity without imposing mechanical resistance or sensory distraction.

In summary, participants' speculative visions motivate investigating how interactive technologies might combine explicit guidance with opportunities for learners to interpret and regulate their own bodily experience.

\section{Discussion}

While singing-centric interaction has drawn growing interest within music technology and HCI~\cite{kilic2023corsetto, cotton2021symbiosinging, cotton2021body, reed2021surface, reed2022singing, chen2026sensing}, interactive systems typically treat respiration either as an explicit controller or as an ambient tool for passive relaxation. Few frameworks have examined breath regulation through the specific interactional tensions identified in our study: the \textit{Opacity of Inner Sensations} and the \textit{Disconnected Regulation of Inner Sensations}. 
Biofeedback can make bodily activity available through sensor-based representations, including cues for monitoring and correction~\cite{schwartz2017biofeedback, sigrist2013augmented}. Our participants' accounts foreground a complementary design concern: how learners interpret such external references in relation to felt sensations and intended adjustments. Their reported difficulties suggest the value of considering ongoing bodily calibration alongside the detection and correction of performance deviations.

To address these tensions, we introduce the \textbf{Somatic Alignment Model} (Figure~\ref{fig:unified-somatic-model}). Rather than functioning as a prescriptive diagnostic tool or an absolute evaluative metric, SAM serves as a generative design lens. Building on soma design's attention to cultivated bodily experience~\cite{schiphorst2011self, hook2018designing}, SAM focuses on how additional perspectives may support the development, negotiation, or reconsideration of somatic understanding when learners find their own sensations difficult to interpret. 
The mismatch themes motivate distinguishing the interpretation of sensation from its use in intended adjustment. Participants' accounts of touch, demonstration, and imagery motivate examining how additional perspectives enter that process. Their requests for explicit cues, alongside concerns about dependence, foreground the distribution of authority. SAM organizes these relationships, while the design space develops questions about representation, authority, and feedback timing.
We first describe SAM's coupled loops, then develop the design space and illustrate its use in system design and first-person inquiry.

\subsection{The Somatic Alignment Model}
\label{subsec:sam-model}

\begin{figure*}[t]
\centering
\includegraphics[width=0.85\linewidth]{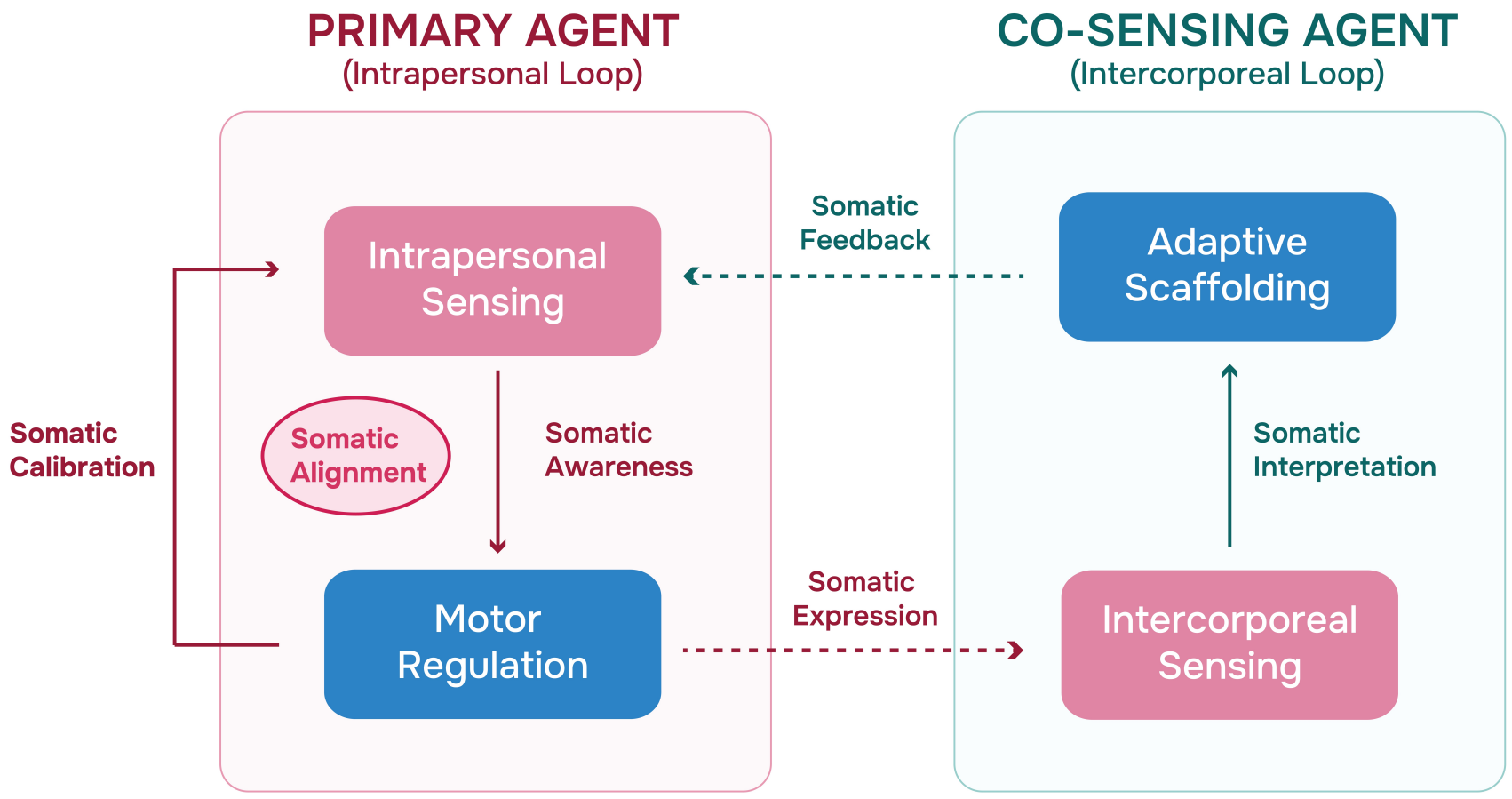}
\caption{\textbf{The Somatic Alignment Model.}
SAM conceptualizes somatic regulation through two coupled processes. Within the \textit{Primary Agent}, an intrapersonal self-sensing loop supports the ongoing calibration of internal sensation and motor regulation. When this process is difficult to navigate from within alone, a \textit{Co-Sensing Agent}---such as an instructor, peer, or adaptive system---can provide an additional bodily perspective through an intercorporeal co-sensing loop. By sensing the \textit{Primary Agent}'s externalized bodily cues, interpreting them contextually, and returning adaptive feedback, the \textit{Co-Sensing Agent} is intended to support the \textit{Primary Agent}'s own somatic calibration.}
\label{fig:unified-somatic-model}
\Description{A block diagram contrasting two coupled feedback loops: the \textit{Primary Agent} on the left and the \textit{Co-Sensing Agent} on the right. The \textit{Primary Agent} frame outlines an intrapersonal loop where a pink box for (1) Intrapersonal Sensing transitions downward via Somatic Awareness to a blue box for (2) Motor Regulation, which loops back upward through Somatic Calibration. An oval labeled Somatic Alignment sits between sensing and regulation. The \textit{Co-Sensing Agent} frame outlines an intercorporeal loop where a pink box for (3) Intercorporeal Sensing leads upward via Somatic Interpretation to a blue box for (4) Adaptive Scaffolding. Connecting the two agents, an arrow labeled Somatic Expression extends from Motor Regulation to Intercorporeal Sensing, while an opposing arrow labeled Somatic Feedback extends from Adaptive Scaffolding back toward Intrapersonal Sensing.}
\end{figure*}

Error-based motor learning models have been particularly effective for actions in which performance can be evaluated against observable task outcomes, such as reaching or pointing~\cite{wolpert_Internal_1995a, wolpert2011principles}. Their direct application becomes less straightforward, however, when the object of regulation is a deep, non-visible bodily process whose relevant sensations are only partially accessible and whose functional baseline varies across contexts~\cite{proske2012proprioceptive, kleber2014neuroscience}. In such cases, effective regulation cannot be understood solely as reducing deviation from a fixed mechanical target. It also depends on continuously relating intended action to felt bodily sensation and physical effort.

The \textit{Somatic Alignment Model (SAM)} formalizes this process through two coupled mechanisms: an \textit{intrapersonal self-sensing loop}, through which a \textit{Primary Agent} senses, regulates, and recalibrates their own bodily experience; and an \textit{intercorporeal co-sensing loop}, through which a \textit{Co-Sensing Agent} observes, interprets, and scaffolds aspects of that experience that may be difficult to discern from within alone (Figure~\ref{fig:unified-somatic-model}). In simplified terms, the intrapersonal loop moves from \textit{sensing} to \textit{awareness}, \textit{regulation}, and \textit{recalibration}, while the intercorporeal loop moves from \textit{observation} to \textit{interpretation}, \textit{scaffolding}, and the return of experiential feedback. The latter is intended to support rather than supersede the former: the purpose of co-sensing is to strengthen the \textit{Primary Agent}'s capacity for autonomous bodily regulation.

SAM is empirically grounded in vocal breath regulation, where practitioners must coordinate respiratory mechanics that are simultaneously consequential for performance and difficult to perceive directly. We therefore do not propose SAM as a universal account of bodily regulation. Rather, we offer it as a generative lens for interactive contexts involving tacit, non-visible, or difficult-to-interpret bodily processes in which people must progressively develop sensitivity to their own internal states.

\subsubsection{The Intrapersonal Self-Sensing Loop: Navigating Internal Opacity}

The intrapersonal loop describes how the \textit{Primary Agent} develops the capacity to perceive and regulate bodily processes from within, drawing on accounts of the lived body~\cite{merleau2013phenomenology} and somaesthetic reflection~\cite{shusterman2008body}. The loop consists of three interdependent processes: sensing and awareness, motor regulation and calibration, and the ongoing emergence of somatic alignment.

\textit{(1) Intrapersonal Sensing to Somatic Awareness.}
The loop begins with \textit{Intrapersonal Sensing}, through which the \textit{Primary Agent} encounters interoceptive and proprioceptive signals arising from their own body~\cite{proske2012proprioceptive}. These signals may initially be vague, ambiguous, or difficult to localize. Through directed attention and repeated bodily exploration, they can become increasingly meaningful as \textit{Somatic Awareness}: an experiential understanding of how particular sensations relate to one's ongoing physical activity. In this sense, awareness is not simply the detection of a physiological signal, but an active process of making sense of bodily experience~\cite{varela2017embodied}.

\textit{(2) Somatic Awareness to Motor Regulation and Calibration.}
Developing awareness enables \textit{Motor Regulation}, in which the \textit{Primary Agent} intentionally adjusts bodily organization---for example, by releasing neck tension or allowing greater expansion around the lower ribs. Yet intended regulation does not necessarily produce the desired bodily effect. Deliberate attempts to `engage' or `support' can instead generate compensatory contraction elsewhere in the body. \textit{Somatic Calibration} therefore closes the intrapersonal loop: the \textit{Primary Agent} attends to the felt consequences of an adjustment and evaluates whether it produces greater functional ease or introduces unintended effort. Regulation is consequently iterative rather than corrective in a single step; each adjustment generates new bodily sensations that become the basis for subsequent sensing and recalibration.

\textit{(3) Somatic Alignment as an Ongoing Process.}
Repeated cycles of sensing, regulation, and calibration can support what we term \textit{Somatic Alignment}. We define \textit{Somatic Alignment} not as a fixed posture, objectively optimal configuration, or final state to be attained, but as an ongoing, situated process through which felt bodily sensation and intended regulation become experientially coherent for the \textit{Primary Agent}. Reduced unwanted effort and greater self-trust are potential outcomes to investigate, rather than effects established by the present study. Because internal bodily conditions fluctuate across situations, alignment must be continually renegotiated rather than permanently achieved.

Phenomenologically, this process can involve the gradual incorporation of initially deliberate regulation into the pre-reflective body schema~\cite{gallagher2005body}. What first requires explicit attention may, through repeated calibration, become increasingly available as an intuitive mode of bodily organization. Within SAM, \textit{Somatic Calibration} therefore describes the mechanism through which the \textit{Primary Agent} continually evaluates and adjusts bodily regulation, whereas \textit{Somatic Alignment} describes the emergent coherence that this process can temporarily sustain.

\subsubsection{The Intercorporeal Co-Sensing Loop: Shared Scaffolding and Feedback}

Although bodily experience remains fundamentally grounded in the first-person perspective~\cite{loke2018somatic}, self-sensing may reach an impasse when sensations remain obscure, familiar patterns become difficult to notice, or the \textit{Primary Agent} lacks an experiential reference for interpreting what they feel. SAM therefore introduces an intercorporeal co-sensing loop in which another person or computational system gains an additional perspective on the \textit{Primary Agent}'s bodily activity without claiming privileged authority over their lived experience~\cite{merleau2013phenomenology, turmo2023intercorporeal}. Here, \textit{Co-Sensing Agent} names a functional role; human partners draw on embodied perception, whereas computational systems operate through sensed data and designed inference rules. This loop proceeds through somatic expression and intercorporeal sensing, contextual interpretation and scaffolding, and the return of somatic feedback.

\textit{(1) Somatic Expression and Intercorporeal Sensing.}
As the \textit{Primary Agent} regulates their body, aspects of an otherwise internal process become available to others through \textit{Somatic Expression}. Respiratory coordination, for example, may manifest through vocal timbre, torso movement, posture, muscular tension, or patterns of expansion that can be perceived through physical contact. These expressions can therefore be sensed directly by another embodied person: an instructor may watch the movement of the ribcage, listen to changes in vocal production, place a hand on the torso to feel respiratory expansion, or use their own body as a perceptual reference during shared movement.

Interactive sensing technologies can extend rather than replace this embodied perceptual relation by rendering physiological traces that may otherwise remain difficult to perceive, such as fine-grained ribcage expansion, muscle activity, or breathing cadence. Through \textit{Intercorporeal Sensing}, the \textit{Co-Sensing Agent} thus attunes to the \textit{Primary Agent}'s bodily expression through one or more channels---visual, auditory, tactile, kinesthetic, or sensor-mediated---to develop an additional perspective on the \textit{Primary Agent}'s ongoing somatic regulation.

\textit{(2) Somatic Interpretation and Adaptive Scaffolding.}
What is perceived through intercorporeal sensing is not self-explanatory, regardless of whether it is encountered directly through another person's body or mediated through sensing technology. The same apparent stillness, muscular firmness, respiratory pattern, or sensor trace may correspond to different bodily experiences depending on the \textit{Primary Agent}'s intention, context, and broader bodily organization. Through \textit{Somatic Interpretation}, the \textit{Co-Sensing Agent} therefore makes situated sense of these perceptual cues rather than treating any single observation or physiological signal as a definitive indicator of correct or incorrect regulation. When external observations and first-person reports diverge, the discrepancy becomes material for further calibration. Neither perspective alone establishes the cause; subsequent inquiry can examine how a change is experienced and how it relates to the intended action.

This interpretation informs \textit{Adaptive Scaffolding}: support adjusted to the \textit{Primary Agent}'s current understanding and response. Drawing on pedagogical accounts of contingency, fading, and transfer of responsibility~\cite{van2010scaffolding}, SAM asks how external guidance can respond to changing needs while supporting the learner's own calibration. Such support may include explicit direction; its role is reconsidered as the learner responds.

\textit{(3) Somatic Feedback as Reflective Inquiry.}
Adaptive scaffolding is returned to the \textit{Primary Agent} through \textit{Somatic Feedback}. Such feedback may itself be embodied and interpersonal, for example, through touch, guided movement, physical demonstration, verbal imagery, or another person's bodily response, or technologically mediated through visual, auditory, tactile, or multimodal representations. Its function within SAM is therefore broader than communicating an error signal. Rather, feedback can operate as an experiential probe~\cite{gaver1999design}: making otherwise difficult-to-notice bodily relationships available for renewed attention and interpretation.

The \textit{Primary Agent} can then compare this feedback with felt experience and use that comparison in subsequent calibration, completing the connection between the two loops.
The intercorporeal loop therefore returns attention to, rather than away from, the intrapersonal loop. In this way, the \textit{Co-Sensing Agent} contributes an additional embodied perspective while preserving the \textit{Primary Agent} as the ultimate site of bodily sensing, interpretation, and regulation.

\subsection{Operationalizing SAM: A Generative Design Space}
\label{subsec:sam-design-space}

\begin{figure*}[t]
\centering
\includegraphics[width=0.9\linewidth]{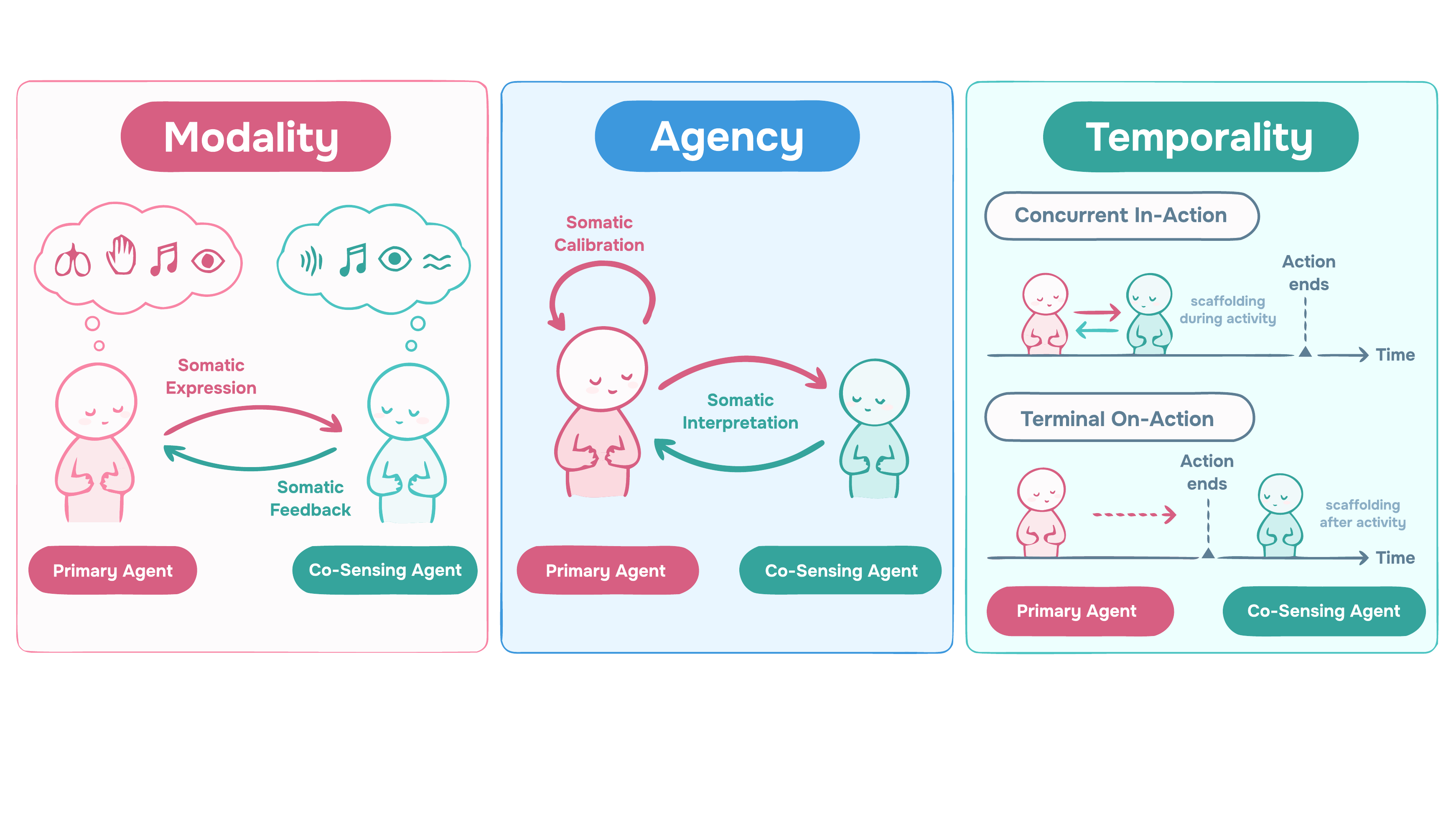}
\caption{\textbf{Three design dimensions of the Somatic Alignment Model.}
\textit{Modality} concerns how bodily information is expressed and returned between agents;
\textit{Agency} concerns how interpretive and regulatory authority is negotiated between the \textit{Primary Agent}'s somatic calibration and the \textit{Co-Sensing Agent}'s interpretation;
\textit{Temporality} concerns whether scaffolding occurs during bodily activity or after it ends.}
\label{fig:sam-design-space}
\Description{Three panels arranged from left to right show Modality, Agency, and Temporality. Primary Agents are pink, and Co-Sensing Agents are teal. In the Modality panel, sensory icons appear above both agents. A pink arrow labeled Somatic Expression points from the Primary Agent to the Co-Sensing Agent, and a teal arrow labeled Somatic Feedback points back. In the Agency panel, the Primary Agent is larger than the Co-Sensing Agent. A pink self-directed arrow above the Primary Agent is labeled Somatic Calibration. Two curved arrows connecting the agents form a loop labeled Somatic Interpretation, representing the relationship between first-person calibration and external interpretation. The Temporality panel contains two horizontal timelines, each with a dashed vertical marker labeled Action ends. In Concurrent In-Action, both agents and their bidirectional exchange appear before this marker, with scaffolding during activity. In Terminal On-Action, the Primary Agent appears before the marker and the Co-Sensing Agent after it, with scaffolding after activity.}
\end{figure*}

We translate SAM into three generative design dimensions: \textbf{Modality}, \textbf{Agency}, and \textbf{Temporality}. As Figure~\ref{fig:sam-design-space} shows, these dimensions correspond to three fundamental properties of co-sensing: \textit{Modality} concerns how bodily information is expressed and returned between agents; \textit{Agency} concerns how interpretive and regulatory authority is distributed between them; and \textit{Temporality} concerns when external scaffolding enters the \textit{Primary Agent}'s ongoing regulation. These dimensions provide generative coordinates rather than an exhaustive taxonomy.

\subsubsection{Modality}

\textit{Modality} concerns how bodily experience becomes perceptible across the boundary between the \textit{Primary Agent} and \textit{Co-Sensing Agent}, both through \textit{Somatic Expression} and through the return of \textit{Somatic Feedback}. These exchanges may occur directly through embodied perception---such as seeing, hearing, touching, or physically feeling another person's movement---or may be extended through sensing and feedback technologies. Designing along this dimension involves negotiating the degree of correspondence between bodily processes and their perceptible representation, which we characterize through two design poles: \textit{Isomorphic Mappings} and \textit{Metaphoric Abstraction}.

\textit{Isomorphic Mappings} preserve a relatively direct correspondence between a bodily process and the form through which it is sensed or returned. In human--human interaction, for example, an instructor may place a hand on the ribcage to directly perceive respiratory expansion or use touch to draw attention to a specific anatomical region. Interactive technologies can similarly monitor localized bodily activity, such as muscle activation or rib movement, and return cues to the same or closely corresponding anatomical location (e.g., localized vibrotactile feedback~\cite{piao2023mappemg}). Such mappings can provide concrete anatomical reference points when bodily sensations are initially difficult to locate or differentiate.

However, greater anatomical correspondence does not necessarily produce greater somatic understanding. Highly localized cues may direct attention toward isolated structures while obscuring how those structures participate in broader patterns of bodily coordination. In some contexts, direct stimulation may also elicit compensatory muscular responses rather than the intended release~\cite{piao2022sensing}. Isomorphic mappings should therefore be understood as one possible perceptual anchor rather than as inherently more accurate or desirable representations of bodily experience.

At the other pole, \textit{Metaphoric Abstraction} loosens direct anatomical correspondence and represents bodily dynamics through more holistic or cross-modal qualities. A human instructor may use verbal imagery, bodily demonstration, or guided movement to communicate qualities such as openness, expansion, continuity, or release. Interactive systems can similarly transform physiological dynamics into acoustic, spatial, visual, tactile, or material metaphors, such as expanding shape-changing fabrics~\cite{kilic2023corsetto} or responsive soundscapes~\cite{hook2018designing, reed2023negotiating}. Rather than identifying a specific anatomical structure, such mappings can function as experiential mirrors that invite the \textit{Primary Agent} to interpret bodily organization as an integrated whole.

This abstraction introduces its own trade-off. Because metaphoric representations intentionally distance themselves from anatomical correspondence, they may provide less spatial specificity when a practitioner needs to identify a particular bodily region or mechanical relationship. Designers therefore need not treat isomorphic and metaphoric approaches as mutually exclusive. They may instead move between them as somatic calibration develops: providing relatively concrete anatomical anchors when orientation is needed, while introducing more abstract representations when broader coordination, exploration, or expressive fluidity becomes more important.

\subsubsection{Agency}

\textit{Agency} concerns the distribution and negotiation of interpretive and regulatory authority between the \textit{Primary Agent} and \textit{Co-Sensing Agent}. Within SAM, both agents can participate in making sense of bodily activity: the \textit{Primary Agent} does so through first-person \textit{Somatic Calibration}, while the \textit{Co-Sensing Agent} contributes an additional perspective through \textit{Somatic Interpretation}. The design question is therefore not simply which agent ``controls'' the interaction, but how much authority is temporarily afforded to each perspective and how that distribution changes over time. We characterize this negotiation through two design orientations: \textit{Prescriptive Direction} and \textit{Liminal Presence}.

\textit{Prescriptive Direction} places greater interpretive and regulatory authority with the \textit{Co-Sensing Agent}. An instructor might explicitly identify a bodily pattern and instruct the \textit{Primary Agent} how to modify it, while an interactive system might evaluate sensed activity against predefined criteria and issue corrective prompts. Such guidance can provide useful reference points when the \textit{Primary Agent} has little basis for interpreting unfamiliar bodily sensations. However, when external correction becomes continuous or overly authoritative, it may shift attention away from the consequences of movement as felt by the \textit{Primary Agent}. Research on augmented motor feedback similarly suggests that excessive or persistent guidance can foster dependence on external information and reduce learning once that guidance is withdrawn~\cite{salmoni1984knowledge, schmidt2018motor}. Within SAM, the concern is therefore not prescription itself, but the possibility that external interpretation begins to substitute for the \textit{Primary Agent}'s own process of somatic calibration.

\textit{Liminal Presence} places greater regulatory initiative with the \textit{Primary Agent} while maintaining the \textit{Co-Sensing Agent} as an available but non-dominant perceptual resource. Drawing on principles of calm and peripheral interaction~\cite{bakker2015peripheral}, the \textit{Co-Sensing Agent} does not continuously classify bodily variation as correct or incorrect. Instead, scaffolding remains at the margins of attention and becomes more salient when an additional perspective may help the \textit{Primary Agent} reconsider an otherwise unnoticed or difficult-to-interpret bodily pattern. In human interaction, this may involve an instructor withholding immediate correction and instead using a light touch, demonstration, or exploratory question. In an interactive system, it may involve subtle peripheral cues rather than persistent threshold alarms. Such an orientation preserves space for variation and exploration while allowing the intrapersonal calibration loop to remain the primary site of regulation.

These orientations should not be understood as a binary choice between external guidance and complete autonomy. Early exploration, unfamiliar sensations, or temporary uncertainty may justify more explicit scaffolding, whereas increasing familiarity may support a gradual redistribution of authority toward the \textit{Primary Agent}. Designers can therefore construct \textit{fading trajectories} in which the frequency, specificity, or assertiveness of external intervention decreases as the \textit{Primary Agent} develops a more stable capacity for self-sensing and calibration. This reduction need not be permanent: changes in bodily condition, task demands, or confidence may make renewed scaffolding useful. Agency, in this sense, concerns how interpretive and regulatory authority is negotiated between the \textit{Primary Agent} and \textit{Co-Sensing Agent} as support needs change, with the aim of sustaining the \textit{Primary Agent}'s own somatic calibration.

\subsubsection{Temporality}

\textit{Temporality} concerns when the \textit{Co-Sensing Agent}'s scaffolding enters the \textit{Primary Agent}'s ongoing cycle of sensing, regulation, and calibration. Feedback may accompany bodily action as it unfolds, or it may be withheld until action has concluded so that the \textit{Primary Agent} can first experience the movement without external interruption. We characterize these alternatives through \textit{Concurrent In-Action Scaffolding} and \textit{Terminal On-Action Scaffolding}.

\textit{Concurrent In-Action Scaffolding} occurs while bodily action is actively unfolding. A human instructor may guide movement through touch, demonstration, verbal imagery, or physical coupling in real time, while an interactive system may continuously translate sensed bodily activity into auditory, visual, or tactile cues. Such temporal coupling can help the \textit{Primary Agent} immediately relate an external perceptual reference to an ongoing bodily sensation. For continuous activities such as respiratory coordination across a musical phrase, this immediacy can support moment-to-moment exploration and adjustment.

At the same time, concurrent scaffolding competes with the sensory and attentional demands of ongoing action. External cues can divide attention or interrupt automatic coordination when they become too salient or information-dense~\cite{sigrist2013augmented}. This tension is particularly relevant in expressive or highly coordinated activity, where continuously monitoring an external representation may shift attention away from both first-person bodily sensation and the activity itself. The design challenge is therefore not simply whether feedback is real-time, but how perceptually demanding that feedback becomes while action is unfolding.

\textit{Terminal On-Action Scaffolding} separates external scaffolding from active execution and introduces it after a movement, exercise, or performance segment has concluded. Drawing on Sch\"{o}n's concept of reflection-on-action~\cite{schon1986reflective}, the \textit{Co-Sensing Agent} can use the completed action as a shared object for subsequent inquiry. A human partner may describe what they saw, heard, or physically felt during the interaction, while a technological system may present previously recorded physiological traces, tactile replays, movement visualizations, or synchronized performance audio. These externalized perspectives can then be compared with the \textit{Primary Agent}'s remembered bodily experience, creating a dedicated space for retrospective \textit{Somatic Calibration}.

Delayed scaffolding preserves greater attentional continuity during execution, but it also introduces distance between the original sensation and its later interpretation. Designers may therefore coordinate temporal strategies rather than adopting either pole exclusively: subtle or peripheral cues can accompany action when immediate orientation is useful, while richer interpretation and reflection can be reserved for pauses or post-action review. Temporality thus determines not only when information is delivered, but how external sensing is woven into the rhythm of first-person bodily experience.

\subsection{Applying SAM: System Design and Methodological Scaffolding}
\label{subsec:subversion}

From this perspective, a \textit{Co-Sensing Agent}, whether a person, an interactive system, or a combination of both, functions as an adaptive scaffold rather than as an unquestioned authority on bodily experience. Its role can expand when an additional perspective is useful and recede when the \textit{Primary Agent} can productively continue through self-sensing. Because \textit{Somatic Alignment} is continually renegotiated rather than permanently attained, this relationship need not follow a simple linear progression from dependence to independence. Instead, the degree of scaffolding may change with context, bodily condition, task demands, and the \textit{Primary Agent}'s momentary capacity for calibration.

Below, we illustrate two ways in which SAM can inform HCI practice: rethinking existing bodily feedback systems and structuring first-person soma design inquiry. The first example uses SAM to examine an existing system's design choices and identify possible alternatives; the second considers how SAM can frame collaborative reflection on bodily experience. Both examples concern music performance, extending the discussion from vocal pedagogy to piano-related systems and inquiry.

\subsubsection{Rethinking Existing Biofeedback Systems}

Many performance-oriented biofeedback systems externalize otherwise difficult-to-perceive physiological processes through real-time sensing and feedback. These representations can be useful, but designs that rely heavily on continuous, localized, or evaluative cues may inadvertently privilege externally measured states over the \textit{Primary Agent}'s developing bodily awareness. SAM provides a vocabulary for examining and reconfiguring these relationships rather than treating biofeedback itself as inherently prescriptive.

Consider an instrumental learning interface that monitors forearm electromyography and maps muscular exertion to localized vibrotactile feedback during piano practice~\cite{piao2023mappemg}. Such a configuration occupies a particular region of the SAM design space: its \textit{Modality} is relatively isomorphic because muscular activity is returned through localized tactile stimulation; its \textit{Temporality} is concurrent because feedback occurs during performance; and, when vibration functions as an explicit indication of excessive exertion, its \textit{Agency} may lean toward \textit{Prescriptive Direction}. This configuration can make otherwise difficult-to-perceive muscular activity salient, but it may also direct attention toward a localized physiological signal and compete with the performer's ongoing sensory and musical activity.

SAM makes it possible to ask how the same sensing capability could support different forms of somatic calibration. Along the \textit{Modality} dimension, raw muscular activity might be retained as a localized anatomical reference when spatial orientation is useful, but alternatively translated into broader auditory, visual, or tactile qualities when the design goal shifts toward whole-body coordination. Along the \textit{Agency} dimension, feedback need not function continuously as a threshold-based correction; the system might instead reduce the frequency or assertiveness of its intervention as the \textit{Primary Agent} becomes more able to recognize corresponding sensations without external prompting. Along the \textit{Temporality} dimension, detailed physiological representations could be deferred until a pause in performance, allowing the performer to compare recorded activity with remembered bodily sensation without continuously attending to an external display.

This example illustrates how the same sensing capability can support different mappings, distributions of authority, and feedback timings.
SAM foregrounds how technological information participates in somatic regulation. Re-architecting the system therefore means designing the relationship between external sensing and first-person calibration, rather than maximizing the amount or immediacy of feedback provided.

\subsubsection{Structuring First-Person Soma Design}

SAM can also function as a methodological heuristic for bodily inquiry within Soma Design~\cite{hook2018designing, kaltenhauser2024playing, bala2023towards, cavdir2026performing}. First-person approaches provide privileged access to lived bodily experience, yet tacit and pre-reflective processes can be difficult to articulate while they are unfolding. Attending analytically to one's own movement may alter the experience itself, while later reflection can lose aspects of sensations that were fleeting or difficult to verbalize.

SAM offers one way to address this tension by structuring first-person inquiry as a scaffolded, multi-perspective process rather than as solitary introspection alone. This possibility is illustrated in the investigation of respiratory dynamics and expressive phrasing in piano performance~\cite{Piao26NIME}. During these inquiry sessions, bodily understanding was distributed across the performer's first-person experience, the interviewer's embodied observations, and physiological traces captured by the interactive prototype. 

The \textit{Agency} and \textit{Temporality} dimensions help articulate how these perspectives were coordinated. During performance, external intervention remained limited so that the performer could remain engaged with musical and bodily experience. After performance, sensor-derived respiratory traces synchronized with audio provided an additional perspective that could be considered alongside the performer's remembered sensations. The interviewer used these externalized traces not as diagnostic evidence of a correct respiratory pattern, but as shared reflective anchors for open-ended somatic inquiry. The traces helped identify moments for comparison with the performer's remembered sensations.
Read through SAM, this configuration suggests a methodological sequence: preserve space for first-person experience during action, then bring recorded traces and another person's observations into retrospective inquiry.

\section{Limitations and Future Work}

While this work offers grounded theoretical and design insights for non-visible somatic interaction, several methodological boundaries and opportunities for future exploration warrant consideration:

\begin{enumerate}

    \item \textbf{Scoping from a Case to Broader Somatic Practices:} Our empirical inquiry intentionally focused on vocal breath control as a \textit{paradigmatic case} of bodily opacity. Because the vocal mechanism operates entirely within the hidden interior of the body, this setting foregrounded reported difficulties in interpreting internal sensations and the pedagogical use of additional bodily perspectives. However, other movement disciplines involve varying balances of internal and external feedback. Domains that engage visible limb trajectories (e.g., dance, martial arts) or physical instruments providing direct mechanical resistance (e.g., acoustic instruments, athletic equipment) offer richer exteroceptive anchors. Future research should explore how SAM translates to these hybrid contexts, examining how internal somatic calibration interfaces with external physical constraints.

    \item \textbf{Methodological Setting and Longitudinal In-Situ Dynamics:} Our two separate workshops elicited learners' and teachers' accounts of bodily difficulties and pedagogical practices. We did not directly observe teacher–learner interactions during instruction, independently measure the relationship between felt sensations and physiological activity, or assess changes in self-regulatory ability. The findings therefore ground a conceptual interpretation of these practices rather than establish their physiological causes or comparative effectiveness. Understanding the nuances of actual embodied learning requires longitudinal observation. Somatic adaptation is a continuous, non-linear process; future work should deploy interactive scaffolds in authentic, longitudinal settings across extended terms to observe how co-sensing dynamics accommodate shifting physiological baselines and evolving motor habits over time.  

\end{enumerate}

Finally, SAM is presented not as a closed, definitive doctrine, but as an open generative framework. Its enduring value to the HCI community will depend on ongoing physical instantiation, empirical deployment, and iterative refinement across diverse bodily and movement practices.

\section{Conclusion}

In this paper, we investigated participants' experiences of perceiving, communicating, and regulating breath in vocal pedagogy as a case of difficult-to-interpret inner bodily sensations. Through reflective workshops and somatic body mapping with vocal learners and experienced pedagogues ($N=9$), we characterized two intrapersonal perception mismatches—the \textit{Opacity of Inner Sensations} and the \textit{Disconnected Regulation of Inner Sensations}—alongside the pedagogical strategies participants described as addressing these difficulties and the interactive interfaces they envisioned.

Translating these empirical insights into interaction principles, we introduced the \textbf{Somatic Alignment Model}, a conceptual framework that couples an \textit{Intrapersonal Self-Sensing Loop} with an \textit{Intercorporeal Co-Sensing Loop}. We operationalized SAM through a three-dimensional generative design space—\textbf{Modality}, \textbf{Agency}, and \textbf{Temporality}—providing an actionable vocabulary for designers to navigate experiential trade-offs between anatomical precision, bodily ease, interpretive authority, and performance flow. Through illustrative applications to biofeedback interfaces and collaborative somaesthetic inquiry, we explored how this space can guide design decisions about the relationship between external feedback and first-person calibration.

Ultimately, this work foregrounds the relationship between external guidance and the cultivation of bodily self-regulation. SAM offers a conceptual lens and generative design vocabulary for exploring how interactive systems might support bodily awareness, personal autonomy, and somatic inquiry. Future studies should examine whether and under what conditions SAM-informed designs support these aims, and how the model transfers beyond vocal pedagogy.

\begin{acks}
This research was supported by an FRQSC Doctoral Scholarship from Fonds de recherche du Québec, the Schulich School of Music at McGill University, and the Centre for Interdisciplinary Research in Music Media and Technology (CIRMMT). The McGill University Research Ethics Board Office approved all data acquired from the focus groups (file number 23-10-079), and participants provided consent. 

We sincerely thank all participants who attended the focus groups and generously contributed their valuable insights. Special thanks to the members of McGill University's Input Devices and Music Interaction Lab (IDMIL) and Music Performance and Body Lab (MPBL) for their ongoing support and feedback throughout the research process. We are particularly grateful to Jacqueline Woodley, Caroline Holden, Colin Raab, Colin Enright, and Mariane Generale for sharing their perspectives as experienced singers. We extend our heartfelt thanks to Professor John Mac Master for his thoughtful suggestions as an excellent voice researcher and opera singer, and to Annika Williams for assisting with participant recruitment and providing helpful input in the early stages of the study. We also deeply appreciate the insights and encouragement from Professors Jun Nishida, Audrey Girouard, and Courtney Reed, who significantly refined our ideas and broadened the scope of this work.

\end{acks}

\bibliographystyle{ACM-Reference-Format}
\bibliography{0-sample-base}

@String{Computing = "Computing" }

@String{Springer = "Springer-Verlag" }

@misc{NVivo,
  author       = {{Lumivero}},
  title        = {NVivo Qualitative Data Analysis Software},
  year         = {2024},
  howpublished = {\url{https://lumivero.com/products/nvivo/}},
  note         = {Accessed: November 27, 2024}
}

@inproceedings{anne2022body,
  title={Body Maps: A generative tool for soma-based design},
  author={Cochrane, Karen Anne and Mah, Kristina and St{\aa}hl, Anna and N{\'u}{\~n}ez-Pacheco, Claudia and Balaam, Madeline and Ahmadpour, Naseem and Loke, Lian},
  booktitle={Proceedings of the Sixteenth International Conference on Tangible, Embedded, and Embodied Interaction},
  pages={1--14},
  year={2022},
  doi={10.1145/3490149.3502262}
}

@book{bauer2013singing,
	author = {Bauer, Karen Tillotson},
	date = {2013},
	isbn = {9780810886896},
	publisher = {Scarecrow Press, Incorporated},
	title = {The Essentials of Beautiful Singing: A Three-Step Kinesthetic Approach},
	year = {2013},
}

@article{krakauer2019motor,
  title={Motor learning},
  author={Krakauer, John W and Hadjiosif, Alkis M and Xu, Jing and Wong, Aaron L and Haith, Adrian M},
  journal={Compr Physiol},
  volume={9},
  number={2},
  pages={613--663},
  year={2019},
  doi={10.1002/cphy.c170043}
}

@article{adolph2020importance,
  title={The importance of motor skills for development},
  author={Adolph, Karen E and Hoch, Justine E},
  journal={Nestl{\'e} Nutrition Institute Workshop Series},
  volume={95},
  pages={136--144},
  year={2020},
  publisher={Karger Publishers},
  doi={10.1159/000511511}
}

@article{palmer2000conceptual,
  title={Conceptual and motor learning in music performance},
  author={Palmer, Caroline and Meyer, Rosalee K},
  journal={Psychological Science},
  volume={11},
  number={1},
  pages={63--68},
  year={2000},
  publisher={SAGE Publications Sage CA: Los Angeles, CA},
  doi={10.1111/1467-9280.00216}
}

@article{proske2018kinesthetic,
  title={Kinesthetic senses},
  author={Proske, Uwe and Gandevia, Simon C},
  journal={Comprehensive physiology},
  volume={8},
  number={3},
  pages={1157--1183},
  year={2018},
  publisher={Wiley Online Library},
  doi={10.1002/cphy.c170036}
}

@inproceedings{piao2023mappemg,
  title={MappEMG: Enhancing music pedagogy by mapping electromyography to multimodal feedback},
  author={Piao, Ziyue and Wanderley, Marcelo M and Verdugo, Felipe},
  booktitle={International Conference on ArtsIT, Interactivity and Game Creation},
  pages={325--341},
  year={2023},
  organization={Springer},
  doi={10.1007/978-3-031-55312-7_24}
}

@incollection{kleber2014neuroscience,
  title={The Neuroscience of Singing},
  author={Kleber, Boris A. and Zarate, Jean Mary},
  booktitle={The Oxford Handbook of Singing},
  editor={Welch, Graham F. and Howard, David M. and Nix, John},
  year={2019},
  publisher={Oxford University Press},
  address={Oxford, UK}
}

@book{leborgne2024vocal,
  title={The Vocal Athlete},
  author={LeBorgne, Wendy D. and Rosenberg, Marci D.},
  year={2024},
  publisher={Plural Publishing},
  address={San Diego, CA, USA}
}

@inproceedings{piao2022sensing,
  title={Sensing the breath: A multimodal singing tutoring interface with breath guidance},
  author={Piao, Ziyue and Xia, Gus},
  booktitle={Proceedings of the International Conference on New Interfaces for Musical Expression},
  address={Auckland, New Zealand},
  year={2022},
  doi={10.21428/92fbeb44.143bb1d4}
}

@inproceedings{reed2021surface,
author = {Reed, Courtney N. and McPherson, Andrew P.},
title = {Surface Electromyography for Sensing Performance Intention and Musical Imagery in Vocalists},
year = {2021},
isbn = {9781450382137},
publisher = {Association for Computing Machinery},
address = {New York, NY, USA},
url = {https://doi.org/10.1145/3430524.3440641},
doi = {10.1145/3430524.3440641},
booktitle = {Proceedings of the Fifteenth International Conference on Tangible, Embedded, and Embodied Interaction},
articleno = {22},
numpages = {11},
location = {Salzburg, Austria},
series = {TEI '21}
}

@article{crocco2021motor,
  title={Motor learning and teaching singing: An overview},
  author={Crocco, Laura and Meyer, David},
  journal={Journal of Singing},
  volume={77},
  number={5},
  year={2021}
}

@book{hixon2006respiratory,
	title = {Respiratory function in singing: {A} primer for singers and singing teachers},
	isbn = {0-9763513-1-5},
	publisher = {Redington Brown},
	author = {Hixon, Thomas J.},
	year = {2006},
}

@inproceedings{cotton2021symbiosinging,
author = {Cotton, Kelsey and Afsar, Ozgun Kilic and Luft, Yoav and Syal, Priyanka and Ben Abdesslem, Fehmi},
title = {SymbioSinging: Robotically transposing singing experience across singing and non-singing bodies},
year = {2021},
isbn = {9781450383769},
publisher = {Association for Computing Machinery},
address = {New York, NY, USA},
url = {https://doi.org/10.1145/3450741.3466718},
doi = {10.1145/3450741.3466718},
booktitle = {Proceedings of the 13th Conference on Creativity and Cognition},
articleno = {52},
numpages = {5},
location = {Virtual Event, Italy},
series = {C\&C '21}
}

@inproceedings{kilic2023corsetto,
author = {Kilic Afsar, Ozgun and Luft, Yoav and Cotton, Kelsey and Stepanova, Ekaterina R. and N\'{u}\~{n}ez-Pacheco, Claudia and Kleinberger, Rebecca and Ben Abdesslem, Fehmi and Ishii, Hiroshi and H\"{o}\"{o}k, Kristina},
title = {Corsetto: A Kinesthetic Garment for Designing, Composing for, and Experiencing an Intersubjective Haptic Voice},
year = {2023},
isbn = {9781450394215},
publisher = {Association for Computing Machinery},
address = {New York, NY, USA},
url = {https://doi.org/10.1145/3544548.3581294},
doi = {10.1145/3544548.3581294},
booktitle = {Proceedings of the 2023 CHI Conference on Human Factors in Computing Systems},
articleno = {181},
numpages = {23},
location = {Hamburg, Germany},
series = {CHI '23}
}

@article{sigrist2013augmented,
  title={Augmented visual, auditory, haptic, and multimodal feedback in motor learning: a review},
  author={Sigrist, Roland and Rauter, Georg and Riener, Robert and Wolf, Peter},
  journal={Psychonomic bulletin \& review},
  volume={20},
  pages={21--53},
  year={2013},
  publisher={Springer},
  doi={10.3758/s13423-012-0333-8},
}

@article{salmoni1984knowledge,
  title={Knowledge of results and motor learning: a review and critical reappraisal.},
  author={Salmoni, Alan W and Schmidt, Richard A and Walter, Charles B},
  journal={Psychological bulletin},
  volume={95},
  number={3},
  pages={355},
  year={1984},
  publisher={American Psychological Association},
  doi={10.1037/0033-2909.95.3.355},
}

@article{zimmerman2012multisensory,
  title={The multisensory brain and its ability to learn music},
  author={Zimmerman, Emily and Lahav, Amir},
  journal={Annals of the New York Academy of Sciences},
  volume={1252},
  number={1},
  pages={179--184},
  year={2012},
  publisher={Wiley Online Library},
  doi={10.1111/j.1749-6632.2012.06455.x}
}

@inproceedings{reed2023negotiating,
  title={Negotiating experience and communicating information through abstract metaphor},
  author={Reed, Courtney N and Strohmeier, Paul and McPherson, Andrew P},
  booktitle={Proceedings of the 2023 CHI Conference on Human Factors in Computing Systems},
  pages={1--16},
  year={2023},
  doi={10.1145/3544548.3580700}
}

@article{wolpert2011principles,
  title={Principles of sensorimotor learning},
  author={Wolpert, Daniel M and Diedrichsen, J{\"o}rn and Flanagan, J Randall},
  journal={Nature reviews neuroscience},
  volume={12},
  number={12},
  pages={739--751},
  year={2011},
  publisher={Nature Publishing Group UK London},
  doi={10.1038/nrn3112}
}

@article{abbruzzese2003sensorimotor,
  title={Sensorimotor integration in movement disorders},
  author={Abbruzzese, Giovanni and Berardelli, Alfredo},
  journal={Movement disorders},
  volume={18},
  number={3},
  pages={231--240},
  year={2003},
  doi={10.1002/mds.10327},
  publisher={Wiley Online Library}
}

@incollection{kawato2007internal,
  title={Internal models for motor control},
  author={Kawato, Mitsuo and Wolpert, Daniel},
  booktitle={Sensory Guidance of Movement: Novartis Foundation Symposium 218},
  pages={291--307},
  year={1998},
  publisher={John Wiley \& Sons},
  address={Chichester, UK},
  doi={10.1002/9780470515563.ch16}
}

@book{schmidt2018motor,
  title={Motor Control and Learning: A Behavioral Emphasis},
  author={Schmidt, Richard A and Lee, Timothy D and Winstein, Carolee and Wulf, Gabriele and Zelaznik, Howard N},
  year={2018},
  publisher={Human kinetics}
}

@article{shadmehr2010error,
  title={Error correction, sensory prediction, and adaptation in motor control},
  author={Shadmehr, Reza and Smith, Maurice A and Krakauer, John W},
  journal={Annual review of neuroscience},
  volume={33},
  number={1},
  pages={89--108},
  year={2010},
  publisher={Annual Reviews},
  doi={10.1146/annurev-neuro-060909-153135}
}

@book{merleau2013phenomenology,
  title={Phenomenology of Perception},
  author={Merleau-Ponty, Maurice},
  note={Translated by Donald A. Landes},
  year={2013},
  publisher={Routledge},
  address={London, UK}
}

@article{turmo2023intercorporeal,
  title={Intercorporeal biofeedback for movement learning},
  author={Turmo Vidal, Laia and M{\'a}rquez Segura, Elena and Waern, Annika},
  journal={ACM Transactions on Computer-Human Interaction},
  volume={30},
  number={3},
  pages={1--40},
  year={2023},
  publisher={ACM New York, NY},
  doi={10.1145/3582428}
}

@phdthesis{reed2023understanding,
	title = {Understanding {Vocal} {Perception}: {Metaphor}-{Based} {Communication} of {Sensory} {Experience}},
	language = {en},
	urldate = {2024-08-27},
	school = {Queen Mary University of London},
	author = {Reed, Courtney N.},
	month = apr,
	year = {2023},
}

@book{hook2018designing,
  title={Designing with the body: Somaesthetic interaction design},
  author={H{\"o}{\"o}k, Kristina},
  year={2018},
  publisher={MIT Press}
}

@inproceedings{cotton2021body,
  title={The Body Electric: A NIME designed through and with the somatic experience of singing},
  author={Cotton, Kelsey and Sanches, Pedro and Tsaknaki, Vasiliki and Karpashevich, Pavel},
  booktitle={Proceedings of the International Conference on New Interfaces for Musical Expression},
  year={2021},
  organization={PubPub},
  doi={10.21428/92fbeb44.ec9f8fdd}
}

@book{miller1996art,
  title={On the Art of Singing},
  author={Miller, Richard},
  year={1996},
  publisher={Oxford University Press}
}

@article{braun2021one,
  title={One size fits all? What counts as quality practice in (reflexive) thematic analysis?},
  author={Braun, Virginia and Clarke, Victoria},
  journal={Qualitative research in psychology},
  volume={18},
  number={3},
  pages={328--352},
  year={2021},
  publisher={Taylor \& Francis},
  doi={10.1080/14780887.2020.1769238}
}

@article{hoch2020historical,
  title={Historical Landmarks in Singing Voice Pedagogy},
  author={Hoch, Matthew},
  journal={Voice and Speech Review},
  volume={13},
  number={1},
  pages={43--60},
  year={2019},
  doi={10.1080/23268263.2018.1527585}
}

@inproceedings{chen2026sensing,
  title={Sensing your vocals: Exploring the activity of vocal cord muscles for pitch assessment using electromyography and ultrasonography},
  author={Chen, Kanyu and Panskus, Rebecca and Wu, Erwin and Peng, Yichen and Saito, Daichi and Kamiyama, Emiko and Li, Ruiteng and Liao, Chen-Chieh and Marky, Karola and Kato, Akira and others},
  booktitle={Proceedings of the 2026 CHI Conference on Human Factors in Computing Systems},
  pages={1--20},
  year={2026},
  doi={10.1145/3772318.3790965}
}

@book{merleau1964primacy,
  title={The Primacy of Perception: And Other Essays on Phenomenological Psychology, the Philosophy of Art, History, and Politics},
  author={Merleau-Ponty, Maurice},
  year={1964},
  publisher={Northwestern University Press}
}

@book{shusterman2008body,
  title={Body Consciousness: A philosophy of Mindfulness and Somaesthetics},
  author={Shusterman, Richard},
  year={2008},
  publisher={Cambridge University Press}
}

@phdthesis{sentianin2018natural,
  title={Natural Singing: A Guide for Using the Alexander Technique to Teach Voice},
  author={Sentianin, Paige},
  year={2018},
  school={University of Colorado}
}

@book{kleinman2013alexander,
  title={The Alexander Technique for Musicians},
  author={Kleinman, Judith and Buckoke, Peter},
  year={2013},
  publisher={Methuen Drama},
  address={London}
}

@book{gelb2013body,
  title={Body Learning: An Introduction to the Alexander Technique},
  author={Gelb, Michael J},
  year={2013},
  publisher={Aurum}
}

@article{goebl2008tactile,
  title={Tactile feedback and timing accuracy in piano performance},
  author={Goebl, Werner and Palmer, Caroline},
  journal={Experimental Brain Research},
  volume={186},
  number={3},
  pages={471--479},
  year={2008},
  publisher={Springer},
  doi={10.1007/s00221-007-1252-1}
}

@incollection{jensenius2010musical,
  title={Musical Gestures: Concepts and Methods in Research},
  author={Jensenius, Alexander Refsum and Wanderley, Marcelo M. and God{\o}y, Rolf Inge and Leman, Marc},
  booktitle={Musical Gestures: Sound, Movement, and Meaning},
  editor={God{\o}y, Rolf Inge and Leman, Marc},
  pages={12--35},
  year={2010},
  publisher={Routledge},
  address={New York, NY, USA}
}

@book{nair1999voice,
  title={Voice Tradition and Technology: A State-of-the-art Studio},
  author={Nair, Garyth},
  year={1999},
  publisher={Singular Publishing Group, Inc.}
}

@article{craig2002you,
  title={How do you feel? Interoception: the sense of the physiological condition of the body},
  author={Craig, Arthur D},
  journal={Nature reviews neuroscience},
  volume={3},
  number={8},
  pages={655--666},
  year={2002},
  publisher={Nature Publishing Group UK London},
  doi={10.1038/nrn894}
}

@article{proske2012proprioceptive,
  title={The proprioceptive senses: their roles in signaling body shape, body position and movement, and muscle force},
  author={Proske, Uwe and Gandevia, Simon C},
  journal={Physiological reviews},
  volume={92},
  number={4},
  pages={1651--1697},
  year={2012},
  publisher={American Physiological Society Bethesda, MD},
  doi={10.1152/physrev.00048.2011}
}

@inproceedings{Stepanova2024intercorporeal,
author = {Stepanova, Ekaterina R. and Reicke, Bernhard E.},
title = {Intercorporeal Design: Dissolving Self-Other Dualism in Interaction Design},
year = {2024},
isbn = {9798400710421},
publisher = {Association for Computing Machinery},
address = {New York, NY, USA},
url = {https://doi.org/10.1145/3686169.3686206},
doi = {10.1145/3686169.3686206},
booktitle = {Proceedings of the Halfway to the Future Symposium},
articleno = {5},
numpages = {9},
location = {Santa Cruz, CA, USA},
series = {HttF '24}
}

@inproceedings{eriksson2019dancing,
  title={Dancing with drones: Crafting novel artistic expressions through intercorporeality},
  author={Eriksson, Sara and Unander-Scharin, {\AA}sa and Trichon, Vincent and Unander-Scharin, Carl and Kjellstr{\"o}m, Hedvig and H{\"o}{\"o}k, Kristina},
  booktitle={Proceedings of the 2019 CHI Conference on Human Factors in Computing Systems},
  pages={1--12},
  year={2019},
  doi={10.1145/3290605.3300847}
}

@inproceedings{patibanda2024exploring,
  title={Exploring shared bodily control: Designing augmented human systems for intra- and inter-corporeality},
  author={Patibanda, Rakesh and Overdevest, Nathalie and Saini, Aryan and Li, Zhuying and Andres, Josh and Knibbe, Jarrod and Van Den Hoven, Elise and Mueller, Florian `Floyd'},
  booktitle={Proceedings of the Augmented Humans International Conference 2024},
  pages={318--323},
  year={2024},
  doi={10.1145/3652920.3653037}
}

@inproceedings{reed2022singing,
  title={Singing knit: soft knit biosensing for augmenting vocal performances},
  author={Reed, Courtney N and Skach, Sophie and Strohmeier, Paul and McPherson, Andrew P},
  booktitle={Proceedings of the Augmented Humans International Conference 2022},
  pages={170--183},
  year={2022},
  doi={10.1145/3519391.3519412}
}

@inproceedings{kaltenhauser2024playing,
  title={Playing with perspectives and unveiling the autoethnographic kaleidoscope in HCI--A literature review of autoethnographies},
  author={Kaltenhauser, Annika and Stefanidi, Evropi and Sch{\"o}ning, Johannes},
  booktitle={Proceedings of the 2024 CHI Conference on Human Factors in Computing Systems},
  pages={1--20},
  year={2024},
  doi={10.1145/3613904.3642355}
}

@inproceedings{bala2023towards,
  title={Towards critical heritage in the wild: Analysing discomfort through collaborative autoethnography},
  author={Bala, Paulo and Sanches, Pedro and Ces{\'a}rio, Vanessa and Le{\~a}o, Sarah and Rodrigues, Catarina and Nunes, Nuno Jardim and Nisi, Valentina},
  booktitle={Proceedings of the 2023 CHI Conference on Human Factors in Computing Systems},
  pages={1--19},
  year={2023},
  doi={10.1145/3544548.3581274}
}

@inproceedings{cavdir2026performing,
  title={Performing Technology: A Five Year Long Autoetnography},
  author={Cavdir, Doga},
  booktitle={Proceedings of the 2026 CHI Conference on Human Factors in Computing Systems},
  year={2026},
  address={Barcelona, Spain},
  publisher={ACM}
}

@INPROCEEDINGS{Piao26NIME,
    address = {London, United Kingdom},
    articleno = {8},
    author = {Ziyue Piao and Yohei Wada and Isabelle Cossette and Marcelo Wanderley and Akira Maezawa},
    booktitle = {Proceedings of the International Conference on New Interfaces for Musical Expression},
    doi = {10.5281/zenodo.20784052},
    issn = {2220-4806},
    month = {June},
    numpages = {10},
    pages = {60--69},
    title = {Exploring Breathing-Music Coupling: Using the Breathing Mirror for Somatic Reflection in Piano Performance},
    track = {Paper},
    url = {http://nime.org/proceedings/2026/nime2026_8.pdf},
    year = {2026}
}

@article{malterud2016sample,
  title={Sample size in qualitative interview studies: guided by information power},
  author={Malterud, Kirsti and Siersma, Volkert Dirk and Guassora, Ann Dorrit},
  journal={Qualitative health research},
  volume={26},
  number={13},
  pages={1753--1760},
  year={2016},
  publisher={Sage Publications Sage CA: Los Angeles, CA},
  doi={10.1177/1049732315617444}
}

@article{wolpert_Internal_1995a,
    title = {An {Internal} {Model} for {Sensorimotor} {Integration}},
    volume = {269},
    issn = {0036-8075, 1095-9203},
    url = {https://www.science.org/doi/10.1126/science.7569931},
    doi = {10.1126/science.7569931},
    language = {en},
    number = {5232},
    urldate = {2025-03-21},
    journal = {Science},
    author = {Wolpert, Daniel M. and Ghahramani, Zoubin and Jordan, Michael I.},
    month = sep,
    year = {1995},
    pages = {1880--1882},
}

@book{varela2017embodied,
  title={The Embodied Mind: Cognitive Science and Human Experience},
  author={Varela, Francisco J and Thompson, Evan and Rosch, Eleanor},
  year={2017},
  publisher={The MIT press}
}

@book{gallagher2005body,
  title={How the Body Shapes the Mind},
  author={Gallagher, Shaun},
  year={2005},
  publisher={Clarendon Press},
  address={Oxford, UK}
}

@article{gaver1999design,
  title={Design: cultural probes},
  author={Gaver, Bill and Dunne, Tony and Pacenti, Elena},
  journal={interactions},
  volume={6},
  number={1},
  pages={21--29},
  year={1999},
  publisher={ACM New York, NY, USA},
  doi={10.1145/291224.291235}
}

@incollection{schiphorst2011self,
  title={Self-evidence: applying somatic connoisseurship to experience design},
  author={Schiphorst, Thecla},
  booktitle={CHI'11 extended abstracts on human factors in computing systems},
  pages={145--160},
  year={2011},
  doi={10.1145/1979742.1979640}
}

@article{loke2018somatic,
  title={The somatic turn in human-computer interaction},
  author={Loke, Lian and Schiphorst, Thecla},
  journal={interactions},
  volume={25},
  number={5},
  pages={54--58},
  year={2018},
  publisher={ACM New York, NY, USA},
  doi={10.1145/3236675}
}

@article{bakker2015peripheral,
  title={Peripheral interaction: characteristics and considerations},
  author={Bakker, Saskia and Van Den Hoven, Elise and Eggen, Berry},
  journal={Personal and ubiquitous computing},
  volume={19},
  number={1},
  pages={239--254},
  year={2015},
  publisher={Springer},
  doi={10.1007/s00779-014-0775-2}
}

@book{schon1986reflective,
  title={The Reflective Practitioner: How Professionals Think in Action},
  author={Sch{\"o}n, Donald A.},
  year={1983},
  publisher={Basic Books},
  address={New York, NY, USA}
}

@book{schwartz2017biofeedback,
  title={Biofeedback: A Practitioner's Guide},
  author={Schwartz, Mark S and Andrasik, Frank},
  year={2017},
  publisher={Guilford Publications}
}

@article{van2010scaffolding,
  title={Scaffolding in teacher--student interaction: A decade of research},
  author={Van de Pol, Janneke and Volman, Monique and Beishuizen, Jos},
  journal={Educational Psychology Review},
  volume={22},
  number={3},
  pages={271--296},
  year={2010},
  publisher={Springer},
  doi={10.1007/s10648-010-9127-6}
}

\end{document}